\documentclass[12pt,letter]{article}
\usepackage{graphicx}
\usepackage{amsmath}
\usepackage{amssymb}
\usepackage{authblk}
\usepackage[left=1in,right=1in,top=1in,bottom=1in]{geometry}
\usepackage{setspace}
\usepackage{float}
\usepackage[super,sort&compress,comma]{natbib}
\usepackage[version=3]{mhchem} 
\usepackage{graphicx} 
\usepackage[font={footnotesize}]{caption} 
\usepackage[labelfont=bf]{caption}
\usepackage{amsmath, amssymb}
\usepackage{natbib}
\usepackage{booktabs}
\usepackage{array}
\usepackage{subcaption}
\usepackage{multirow}
\usepackage{soul}
\usepackage{float}
\usepackage{siunitx}
\usepackage{tabularx}
\usepackage{ulem}

\begin{document}

\title{Determining the Role of Electrostatics in the Making and Breaking of the Caprin1-ATP Nanocondensate}

\author{Maria Tsanai$^{1}$ and Teresa Head-Gordon*$^{1-3}$}
 \date{}
\maketitle
\begin{center}
\vspace{-10mm}
$^1$Kenneth S. Pitzer Theory Center and Department of Chemistry, $^2$Departments of Bioengineering and $^3$Chemical and Biomolecular Engineering, University of California, Berkeley, Berkeley, CA, 94720 USA\\ 

corresponding author: thg@berkeley.edu
\end{center}

\noindent
\textbf{Abstract}
We employ a multiscale computational approach to investigate the condensation process of the C-terminal low-complexity region of the Caprin1 protein as a function of increasing ATP concentration for three states: the initial mixed state, nanocondensate formation, and the dissolution of the droplet as it reenters the mixed state. We show that upon condensation ATP assembles via pi-pi interactions, resulting in the formation of a large cluster of stacked ATP molecules stabilized by sodium counterions. The surface of the ATP assembly interacts with the arginine-rich regions of the Caprin1 protein, particularly with its N-terminus, to promote the complete phase-separated droplet on a lengthscale of tens of nanometers. In order to understand droplet stability, we analyze the near-surface electrostatic potential (NS-ESP) of Caprin1 and estimate the zeta potential of the Caprin1-ATP assemblies. We predict a positive NS-ESP at the Caprin1 surface for low ATP concentrations that defines the early mixed state, in excellent agreement with the NS-ESP obtained from NMR experiments using paramagnetic resonance enhancement. By contrast, the NS-ESP of Caprin1 at the surface of the nanocondensate at moderate levels of ATP is highly negative compared to the mixed state, and estimates of a large zeta potential outside the highly dense region of charge further explains the remarkable stability of this phase separated droplet assembly. As ATP concentrations rise further, the strong electrostatic forces needed for nanocondensate stability are replaced by weaker Caprin1-ATP interactions that drive the reentry into the mixed state that exhibits a much lower zeta potential.

\vspace{-3mm}

\maketitle

\newpage
\section{Introduction}
In biological cells, adenosine triphosphate (ATP) functions as an essential energy source to drive biological reactions such as the synthesis of DNA and RNA \cite{Chu2022}, and for its regulatory role in phosphorylation whereby a phosphate group is added to proteins through ATP hydrolysis \cite{Ren2022, Aida2022}. However it has been noted that the physiological cellular concentration of ATP is in the millimolar range, significantly higher than the micromolar levels needed solely for energy related functions \cite{Mehringer2021, Song2021, Traut1994}. It has been suggested that high ATP concentrations in the cell can help maintain protein solubility, in which ATP acts as a biological hydrotrope, preventing the accumulation of harmful solid protein aggregates, and dissolving them once they form \cite{Patel2017, Sridharan2019, Eastoe2011, Song2021}. Kang et al. further proposed that high ATP concentrations would support a functional role in the process of liquid-liquid phase separation (LLPS) \cite{Kang2018} as demonstrated by proteins such as purified fused in sarcoma (FUS), lysozyme, TDP-43 prion-like domains and Caprin1 \cite{Hu2022, Ren2022, Zalar2023, Aida2022, Kota2024, Liu2023}. 

Typically proteins that undergo LLPS are influenced by their multivalent interactions in driving the phase separation process\cite{Wright2015, Boeynaems2018}, including electrostatics \cite{Brady2017}, cation-pi \cite{Qamar2018, Wang2018}, pi-pi interactions \cite{Vernon2018, Carter-Fenk2023}, and hydrogen bonding \cite{Murthy2019}. Correspondingly, the amphipathic nature of the ATP molecule can simultaneously facilitate a similar variety of molecular interactions with proteins due to its unique chemical structure consisting of a hydrophobic aromatic pyrimidine base, a ribose sugar, and an ionizable triphosphate group that often complexes with metal divalent cations such as magnesium and calcium \cite{Wilson1991}. ATP’s negative charge amplifies its ability to bind to multiple positively charged residues by forming salt bridges \cite{Kota2024, Hu2022, Ou2021}, and when these strong ATP interactions accumulate around the surface of proteins can restrict protein diffusion and thereby promote phase separation \cite{Patel2017, Nishizawa2021, Liu2023, Zalar2023, Choi2024}. However, when ATP becomes oversaturated, it disrupts various electrostatic, hydrophilic, hydrophobic and backbone protein-protein interactions, leading the proteins to reenter a mixed-state \cite{Liu2023, Ren2022, Aida2022,Liu2023}, that is also mediated by the multivalent character of ATP. Furthermore, the hydrophobic nature of the pyrimidine base allows ATP molecules to cluster together via pi-pi interactions formed between adenine groups, and metal cations aid in the formation of ATP clusters by simultaneously interacting with the triphosphate groups of adjacent ATP molecules, acting as cationic linkers to connect them \cite{Ou2021, Sigel2005}. Nishizawa and co-workers using various experimental techniques including NMR, and supported by all-atom molecular dynamics (AA-MD) simulations, found that Mg$^{+2}$-free ATP molecules can spontaneously form dimers or small oligomers even in the absence of proteins in solution \cite{Nishizawa2021}.  

Among all molecular interactions, electrostatic interactions play an especially important role in regulating the LLPS process.\cite{Banani2017, Gomes2019, Tsanai2021, Brasnett2024, Posey2024} Recently introduced NMR paramagnetic relaxation enhancement (PRE) experiments use positively and negatively charged paramagnetic co-solutes to induce changes in $^1$H nuclear magnetizations of protein surface residues, in order to determine the near surface electrostatic potential (NS-ESP) as illustrated by Ishiwara and co-workers \cite{Yu2021}. The NMR-PRE experiments were performed by Toyama et al. in order to measure the residue specific NS-ESP of the C-terminal low-complexity region of the Caprin1 protein in the early mixed state at low ATP concentration, for a large ($>$1 $\mu$m) condensate at moderate levels of ATP and created by fusion of many microscopic droplets driven by centrifugation, and reversion to a mixed state at high ATP concentrations \cite{Toyama2022}. Moreover, Lin et al. combining theoretical and computational methods in the same system showed that interchain ion bridges enhance phase separation, while ATP acts both as a salt-like ion and an amphiphilic hydrotrope \cite{Lin2024}. Welsh and co-workers have shown that higher surface charge density of a microdroplet gives rise to a sufficiently large zeta potential that explains the stability of the phase separated microdroplet state for FUS and PR25:PolyU biological condensates \cite{Welsh2022}. Krainer et al found that LLPS stabilized states of FUS and other proteins can then at reenter a mixed state at high ionic strengths, driven by hydrophobic and non-ionic interactions.\cite{Krainer2021}

Recently it has been noted that there is a size distribution of biological condensates \cite{Martin2021,JFK2022}, ranging from micron-sized in vitro condensates to nanometer scales in the cellular cytoplasm \cite{Cho2018,Rey2020,Keber2024}. It is increasingly appreciated that the differences in condensate size (nanometer vs micron scales) can result in changes in responses to buffers or other constituents \cite{Kar2022,Kar2024}, enzymatic reactivity \cite{GilGarcia2024}, and aggregation, crystallization, and/or amyloid disease mechanisms \cite{Ray2023,Toledo2023}. However, only a few super-resolution microscopy studies\cite{Jain2016,Cho2018,Folkmann2021}, and early molecular dynamics simulations of elastin \cite{Rauscher2017} have characterized how the nanocondensates are spatially organized, and even less is known about why they are stable.  

\begin{figure}[h!]
\centering
\includegraphics[width=0.9\textwidth]{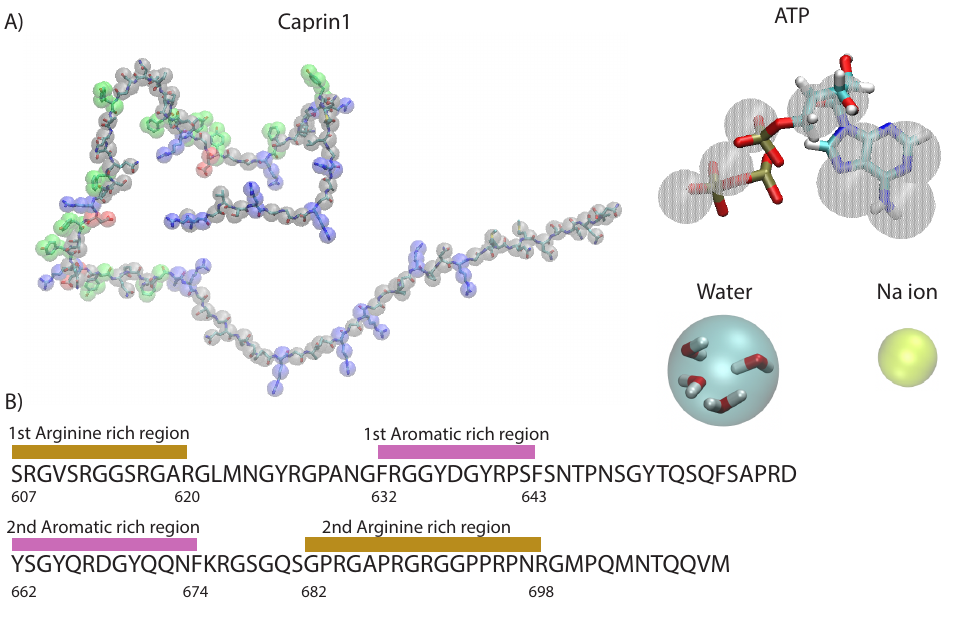}
\caption{\textit{Coarse-grained representations of the molecules and Caprin1 sequence used in this work.} \textbf{(A)} CG representation of the low-complexity C-terminal region of the Caprin1 protein (103 residues), with each type of residue shown in a different color. Mg$^{+2}$-free ATP molecule is shown in gray, water in cyan and sodium counterions in yellow. \textbf{(B)} Amino-acid sequence of Caprin1 protein. The two arginine- and aromatic-rich regions are indicated in orange and pink, respectively along with their residue numbers.} 
 \label{fig:system}
 \vspace {-3mm}
\end{figure}

In this work we employ the coarse-grained (CG) MARTINI model \cite{Souza2021, Sami2023, Pezeshkian2023, Vainikka2023, Marrink2023} to access the time and length scales required for the study of the formation and dissolution of small biomolecular nanocondensates. In particular, we characterize the phase separation process of the Caprin1 protein over a range of 0 to 100 mM ATP concentration (Figure \ref{fig:system}). At moderate ATP concentrations of 10 mM our simulations show the spontaneous formation of a 10-20 nm droplet, with a core composed of tightly stacked ATP molecules stabilized by pi-pi interactions and sodium counterions. This solid ATP core is externally coated by the positively charged Caprin1 proteins, but with a preference for the N-terminal region of Caprin1 absorbed onto the surface of the ATP assembly. By backmapping the mixed states and nanocondensate generated at the CG level back to all-atom (AA) resolution, we can quantify the NS-ESP of Caprin1 and estimate the zeta potential of the nanodroplet, both in the absence and presence of ATP molecules. We find excellent agreement with the NMR-PRE experiments\cite{Toyama2022} for the Caprin1 mixed state at low ATP concentrations, determining a positive NS-ESP of $\sim$20-30 mV. But the NS-ESP of the Caprin1 proteins in the microscopic condensate is highly negative, and the ESP evaluated outside the region of immobile charge provides an estimate of the zeta potential that is indicative of colloidal-like electrostatic forces that explains the remarkable stability of the Caprin1-ATP nanodroplet assembly. Further increases in ATP concentration ultimately dissolves the nanocondensate, attributable to an increased solubilization of the Caprin1 in the aqueous phase, and a transition to weaker cation-pi and pi-pi interactions between Caprin1 and ATP, thereby reducing the NS-ESP and zeta potential to smaller values that explain the Caprin1 reentry into a mixed state.

\section{Methods}

\subsection*{Simulated System}
Explicit-solvent coarse grained molecular dynamics simulations (CG-MD) were conducted with the Martini 3.0 force field \cite{Souza2021} using the GROMACS 2021.5 software \cite{Gromacs2015}. Although the Martini model has been shown to reproduce the conformation of disordered regions \cite{Tsanai2021, Ingolfsson2023}, the protein-water Lennard-Jones interactions were increased by a factor of $\lambda$=1.1 as have been shown by Thomasen et al. to improve the accuracy of the Martini model for disordered proteins \cite{Thomasen2022}. 
The low-complexity C-terminal region of the Caprin1 protein (103 residues) was generated using the martinize.py \cite{Kroon2023} script included in the Martini 3.0 model using a coil secondary structure backbone. The C- and N- terminal beads of the Caprin1 protein have a full $\pm$1e charge. The Martini 3.0 parameters for the Mg$^{+2}$-free ATP molecule were developed and provided by Manuel N. Melo's research group \cite{MeloLab_github}.

A periodic simulation box of 30 $\times$ 30 $\times$ 60 nm$^{3}$ was used for all the simulations and the Caprin1 proteins (10 chains) were randomly dispersed in the box using the GROMACS tool gmx insert-molecules along with the required number of the ATP molecules to achieve the respective ATP concentration. The systems were solvated using normal Martini water beads with gmx solvate and were neutralized by replacing water beads with positively and negatively charged TQ5 beads to represent the Na$^{+}$ and Cl$^{-}$ ion particles. The C-terminal low-complexity region of the Caprin1 protein that was used for all the simulations has a total charge of ${+}$13e, while the Mg$^{+2}$-free ATP molecule has a total charge of ${-}$4e. The steepest descent minimization was followed by a 10 ns equilibration run using a 10 fs time step and the Berendsen barostat with 1 bar and 298 K pressure and temperature, respectively. The equilibration runs were followed by 30 $\mu$s production runs using a 20 fs time step in the NPT ensemble, keeping temperature at 298 K using the v-rescale thermostat \cite{parinello} ($\tau_{T}$ = 1.0 ps$^{-1}$) and pressure at 1.0 bar using the isotropic Parrinello-Rahman pressure coupling \cite{parinelloRahman}  ($\tau_{P}$ = 12 ps$^{-1}$). In accordance with the typical settings associated with the Martini force field, electrostatic interactions were treated using a reaction field approach ($\varepsilon_{rf}$ = 15 beyond the 1.1 nm cut-off) and a shifted Lennard-Jones potential, cut-off at 1.1 nm with the Verlet cut-off scheme \cite{martiniCutoff}.
The neighbor list was updated every 20 steps and coordinates were saved every 200 ps.

Similar procedure was followed for an additional series of CG-MD simulations for systems of only Mg$^{+2}$-free ATP molecules in different concentrations. Here we used a simulation box of 17 $\times$ 17 $\times$ 17 nm$^{3}$, the systems were solvated using normal Martini water beads and were neutralized with Na$^{+}$ ions. After the completion of minimization and equilibration runs, 10 $\mu$s production runs were conducted.

\subsection*{Analysis details}
Snapshots of the Caprin1-ATP systems were obtained using the VMD software \cite{VMD}. Normalized radial distribution functions were computed using the GROMACS tool gmx rdf \cite{Gromacs2015}. This was done by averaging distances from each Caprin1 backbone bead to all other Caprin1 backbone beads not belonging to the same Caprin1 chain, over the final 1 $\mu$s of the trajectory. Similarly the ATP-ATP RDFs were computed by averaging the distances between the center of mass of each ATP molecule and the center of mass of all other ATP molecules excluding itself. The density profiles of the condensates components were computed using the gmx density tool \cite{Gromacs2015} and were normalized by the maximum density value of the specific component. Also the -center flag was used to center the Caprin1 chains in the center of the simulation box.

The number of contacts between each Caprin1 chain and the ATP molecules was calculated using the MDAnalysis python library \cite{MDAnalysis2011, MDAnalysis2016}. A contact was defined when a backbone CG bead from each Caprin1 residue and a CG bead from an ATP molecule were within a cutoff distance of 0.15 nm. These contacts were then averaged across all Caprin1 chains in the system and over ten configurations, taken at 10 ns intervals at the end of the 30 $\mu$s trajectory. The same procedure was followed to calculate the contacts of the Caprin1 chains with Na counterions and water molecules. Caprin1-Caprin1 contact maps were calculated in a similar way, but in this case excluding contacts between residues within the same Caprin1 chain. The contact map is presented as a two-dimensional matrix, with rows and columns representing Caprin1's residues. The details of the diffusion coefficient calculation is discussed in the work of Tsanai and co-workers\cite{Tsanai2021}.

The CG configurations were converted back to their corresponding AA structures using the backward.py program \cite{Wassenaar2014}. Here we use the 2022 updated version of CHARMM36 atomistic force field \cite{Huang2013}. The backward python tool requires mapping files of each molecule, which define the initial positioning of the atoms according to the positions of the CG beads. These mapping files are provided for all the amino-acids, water molecules and Na$^{+}$ ions by the Martini 3.0 model, while for the ATP molecule, we developed them separately using a similar format. With the initram.sh script a series of minimization steps and a position-restrained AA-MD run was performed in order to relax the initial atomistic structure that was generated \cite{Wassenaar2014}.

\subsection*{Electrostatic potential calculations}
Yu et al. \cite{Yu2021} recently developed a protocol for computing the NS-ESP from both Adaptive Poisson-Boltzmann Solver (APBS) \cite{Baker2001,Jurrus2018} calculations, and later extended this approach to AA molecular dynamics simulations \cite{Chen2022}, for a single ubiquitin protein at various KCl concentrations. The protocol optimized the placing of grid points on which the probe molecules sample in order to reproduce the experimental NS-ESP of the ubiquitin protein, and enabled a detailed understanding of the impact of various solution components on its electrostatic profile \cite{Yu2021}.

Making use of the configurations converted to all-atom resolution where each atom holds a partial charge, we calculated the NS-ESP of each residue of the Caprin1 protein. Similarly to the model developed by Yu et al. we placed 3D grid points with a grid spacing of 0.5 {\AA} and for each grid point, a probe with a radius 3.5 {\AA} was used \cite{Yu2021}. To prevent grid points from being placed too close to the atoms, we also accounted for their van der Waals radii. The ESP values for each $^{1}$H nucleus of each Caprin1 backbone residue were calculated based on the contributions of the partial charges from all the other Caprin1 proteins, sodium counterions and Mg$^{+2}$-free ATP molecules using Eq. (1). The contributions of the water molecules where not included in the calculated NS-ESP values, and instead, the solvent's dielectric constant was set to 78. To ensure consistency with the experimental conditions reported by Toyama et al., the charge of the Mg$^{+2}$-free ATP molecules was set to ${-}$3e \cite{Toyama2022} and the NS-ESP values were scaled down by a factor of 1.8.
For each residue in the protein, the near-surface zone  was defined as the region of grid points surrounding the residue's backbone $^{1}$H nuclei, within the distance range that corresponds to a 68\% contribution to: {\[ \sum r_i^{-6} \]} where $r_{i}$ is the distance between the $^{1}$H nucleus and a grid point i \cite{Yu2021}. The pairwise distances of the grid point i with the other atoms present in the system were calculated using the MDAnalysis python library \cite{MDAnalysis2011, MDAnalysis2016} and were averaged across all Caprin1 chains in the system and over ten configurations, taken at 10 ns intervals at the end of the 30 $\mu$s trajectory.

\section{Results} 
We investigated the ATP dependent condensation of the low-complexity C-terminal region of the Caprin1 protein (103 residues) by setting up simulation boxes with 10 Caprin1 proteins in different ATP concentrations of 0.8, 2, 5, 10, and 100 mM in order to cover the same experimental range of ATP that correspond to the initial mixed state, the the nanodroplet formation, and the reentry into the mixed state \cite{Toyama2022}. At physiological pH an individual Caprin1 chain has a net charge of +13, while we used the form of a Mg$^{+2}$-free ATP molecule, which has a net charge of -4, and thus all aqueous ATP-Caprin1 systems were neutralized with the necessary number of sodium counterions. More detail is provided in Methods.

\begin{figure}[h!]
\centering
\includegraphics[width=0.825\textwidth]{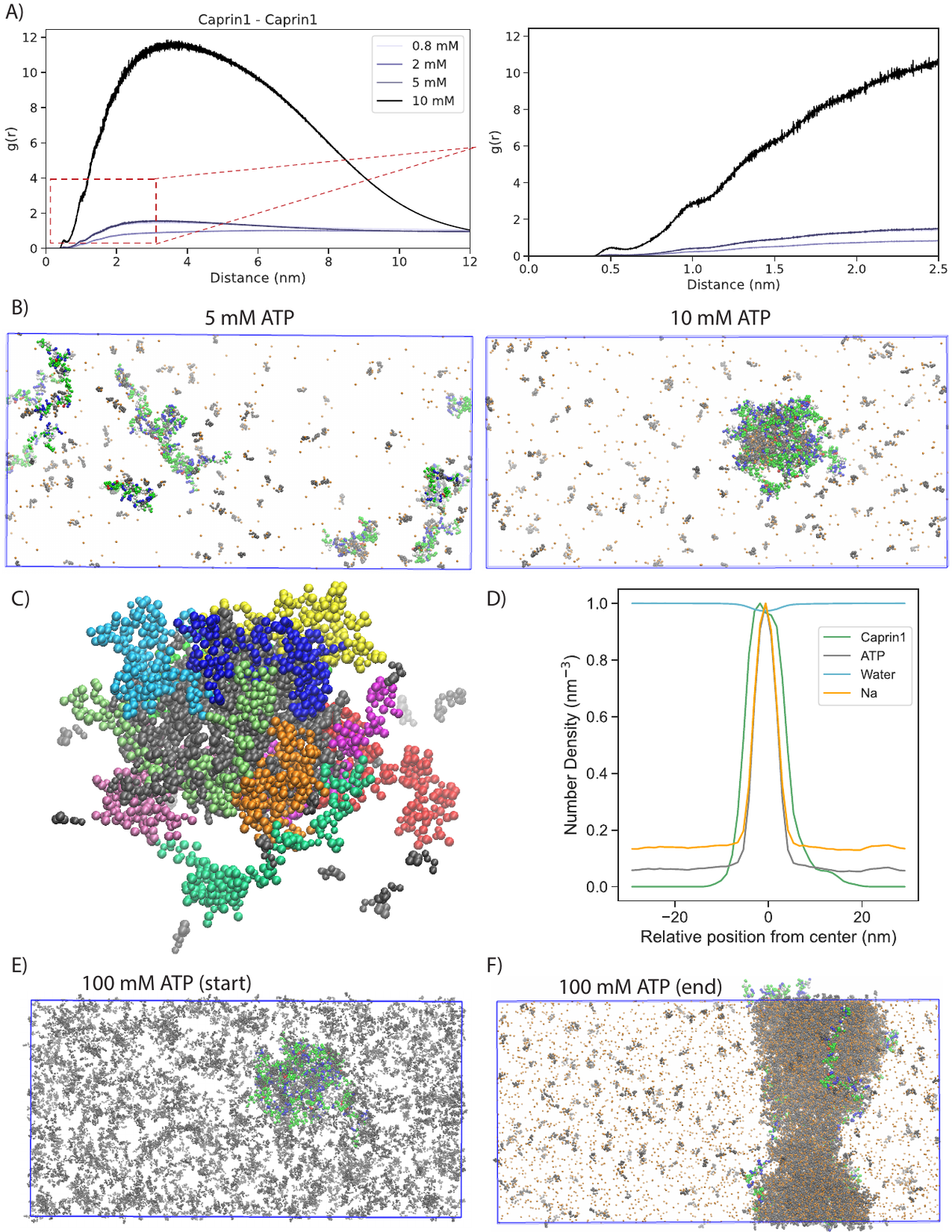}
\caption{\textit{ATP-dependent condensation and dissolution of Caprin1 proteins.} \textbf{(A)} Caprin1-Caprin1 RDFs of the systems and a zoomed-in view showing the smaller distances and the peak at 1 nm. \textbf{(B)} Snapshots of the final configurations of the 30 $\mu$s simulations of 10 Caprin1 chains at different ATP concentrations; water is not shown for clarity. \textbf{(C)} A zoomed-in view of the condensate formed at the 10 mM ATP concentration. Each Caprin1 chain is pictured in a different color and the ATP molecules are pictured in gray. \textbf{(D)} Density profiles of the 10 mM ATP system components, shown in green for the Caprin1 chains, in gray for ATP, in cyan for water molecules and in orange for Na counterions. \textbf{(E)} Initial conditions of the 100mM ATP concentration in which the condensate is formed. \textbf{(F)} After 30 $\mu$s simulations the increase in ATP concentration disrupts the condensate, leading to all Caprin1 proteins transitioning to a mixed state.}
\label{fig:configurations}
\end{figure}

\vspace{-3mm}

\subsection*{Condensation and dissolution of Caprin1 proteins with ATP} To determine the ATP concentration that transitions between no-condensation to condensation in the CG model, we performed 30 $\mu$s simulations starting from randomly distributed Caprin1 proteins and ATP molecules in the simulation box. Figure \ref{fig:configurations}A displays the  computed Caprin1-Caprin1 radial distribution functions (RDFs) at different ATP concentrations ranging from 0.8 to 10 mM. It is apparent that no condensation is observed for any system up through 5 mM ATP concentration as the RDFs display overall uniform Caprin1 densities, whereas at 10 mM ATP concentration there is a dramatic increase in local density over the 2-6 nm length scale. Snapshots from the final configurations of the CG simulations at the different ATP concentrations are shown in Figure \ref{fig:configurations}B. All Caprin1 chains are found to be dispersed in the system at 0.8 mM ATP concentration, while for the 2 and 5 mM ATP concentrations there is formation of small diffuse clusters of 2-3 Caprin1 chains. 

However, upon increasing the ATP concentration to 10 mM, all Caprin1 chains phase separate, giving rise to a distinct and well-defined condensate as seen in Figure \ref{fig:configurations}C and consistent with the onset of large density changes found in the RDFs. The Caprin1 condensate observed in the phase-separated state displays a distinctive spherical shape, with a notable abundance of ATP molecules that are centralized within the inner core of the assembly while the Caprin1 chains are positioned on the outer layer as shown in Figure \ref{fig:configurations}D. We further examined the liquid nature of the condensate that is formed by computing the diffusion coefficients for the water molecules in the supernatant and in the condensate phases to be D$_{supern}$  = (2.02 $\pm$ 0.20) $\times$ 10$^{-9}$ m$^{2}$/s and D$_{cond}$ = (0.23  $\pm$ 0.04) $\times$ 10$^{-9}$ m$^{2}$/s respectively. This difference of one order of magnitude in the diffusion coefficients in the two phases is in agreement with findings for various condensate systems \cite{Kausik2010, Tsanai2021}. Finally, to capture the dissolution of the condensate, we increased the ATP concentration to 100 mM in the system where the condensate had initially formed at 10 mM (Figure \ref{fig:configurations}E) and extended the simulation time for an additional 30 $\mu$s. This significant increase in ATP concentration disrupted the condensate, leading to all Caprin1 proteins transitioning into a mixed state as seen in Figure \ref{fig:configurations}F.  
 
In order to determine whether the extended ATP oligomerization we observe in condensate formation arises from artifactually strong interactions between ATP molecules in the Martini 3.0 model, we performed an additional set of CG-MD simulations containing only ATP molecules at varying concentrations ranging from 0.8 to 10 mM and neutralized with sodium counterions. No oligomers formed in the systems with ATP concentrations up to and including 5 mM, and at most dimers were observed at 10 mM ATP concentration (Supplementary Fig. S1). This is in agreement with the NMR experiments and AA-MD simulations conducted by Nishizawa et al., where only ATP dimers and up to pentamers were identified in systems with high Mg$^{+2}$-free ATP concentrations \cite{Nishizawa2021}. Consequently, the amphipathic nature of ATP allows it to form a phase separated state that simultaneously promotes electrostatic interactions with the positively charged Caprin1 chains, while engaging in pi-pi interactions with each other, forming large stacked assemblies that are further stabilized by sodium counterions. This is consistent with ATP and the Na+ counterions forming a strongly correlated "ionic plasma" \cite{Mondal2025}, but ATP is also an unusual salt whereby LLPS may also be driven by non-ionic effects such as pi-stacking and beyond just Debye-Huckel correlations.

\begin{figure}[h]
\centering
\includegraphics[width=0.9\textwidth]{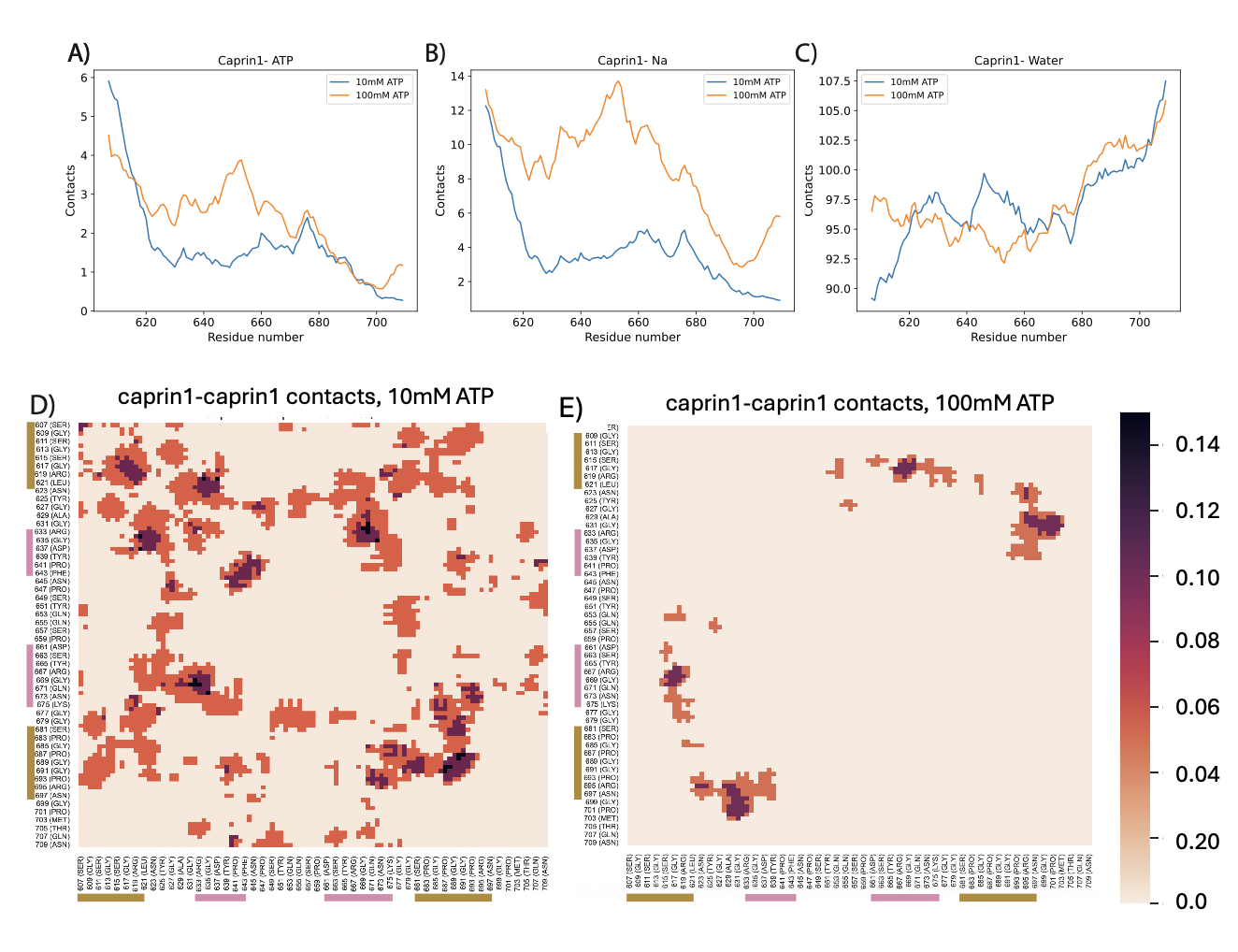}
\caption{\textit{Contact analysis of Caprin1 with the condensate and reentrant state components.} \textbf{(A)} Caprin1-ATP, \textbf{(B)} Caprin1-NA counterions, and \textbf{(C)} Caprin1-water contacts. The contact map between Caprin1-Caprin1 chains for \textbf{(D)} 10 mM ATP corresponding to the condensate and \textbf{(E)} 100 mM ATP corresponding to the reentrant state. The two arginine and aromatic rich regions are indicated in orange and pink, respectively. The frequency of contacts are averaged over all Caprin1 chains and over multiple configurations at the end of the 30 $\mu$s trajectory.}
\label{fig:contacts}
\end{figure}

\vspace{-3mm}

\subsection*{Structural analysis of the Caprin1-ATP condensate and reentry}
Next, we consider the intermolecular contacts made between the Caprin1 chains and the other components of the system when the 15-20 nm droplet is formed and after its dissolution as shown in Figure \ref{fig:contacts}. Interestingly, our analysis shows that each Caprin1 chain is in close contact with an average of 6 ATP molecules, which is consistent with experimental findings of protein contacts made with an average of 5 ATP molecules for each Caprin1 chain \cite{Toyama2022}. Notably for the system at 10mM ATP where the droplet is formed (in blue), the highest amount of contacts for both ATP and Na occur with the 1st arginine rich region (residues 607-620) of Caprin1 of the N-terminus as is evident from Figure \ref{fig:contacts}A,B, whereas the C-terminal region of each Caprin1 chain is oriented towards the bulk water phase for which it exhibits the highest number of interactions with water molecules, as shown in Figure \ref{fig:contacts}C. While both the N- and C- terminal regions are rich in arginine residues, this analysis reveals a significant distinction in their molecular interactions, and is also consistent with the experimental findings, where mutagenesis and residue-specific chemical shifts of Caprin1 emphasized the crucial role of the N-terminal arginine-rich region in driving phase separation, compared to the other arginine regions in the sequence \cite{Wong2020}. Moreover, the contact map between the Caprin1 chains in Figure \ref{fig:contacts}D, reveals an increased number of interactions between the arginine-arginine, aromatic-aromatic, and arginine-aromatic rich regions. These results are also consistent with the experimental NMR findings, which also report intermolecular contacts within these regions \cite{Wong2020, Toyama2022}.

An additional contact analysis was conducted on the system where the condensate is disrupted at 100 mM ATP. Figure \ref{fig:contacts}E shows that in the reentry to the mixed state the Caprin1-Caprin1 contacts have decreased significantly, reflecting the disruption of the condensate. The original protein-protein contacts are seen to be replaced by an increase in the interactions of Caprin1 with ATP molecules and Na counterions, particularly in the protein's two aromatic-rich regions (Figure \ref{fig:contacts}A,B shown in orange) that indicates a transition to pi-pi and cation-pi type interactions. Moreover, the N-terminus of Caprin1 shows increased contacts with water molecules (Figure \ref{fig:contacts}C), in addition to the C-terminus, indicating that the whole Caprin1 protein is more soluble in water as would be expected in the mixed state. 

\subsection*{Electrostatic potentials of Caprin1 as a function ATP concentration}
Thus far we have demonstrated that Caprin1 undergoes self assembly via clustering mediated by ATP to increase the local concentration of the protein, the standard definition of biological condensation \cite{Banani2017}. Further experimental validation of the phase separation of Caprin1-ATP observed in the CG simulation, such as its occurrence  within the concentration range of 5-10 mM Mg$^{+2}$-free ATP, is in good qualitative agreement with the  experimental range of 0.8-2 mM ATP-Mg$^{+2}$ concentrations \cite{Toyama2022}. Many other aspects of the condensate protein-protein and protein-ATP interactions determined from our CG simulation are also in good agreement with experimental observations as described in the previous section. 

However, we wish to provide further support for our simulated observations for the early mixed state, condensate formation, and reentry into the mixed state-- driven by increasing ATP concentration-- through evaluation of the NS-ESP. Recently NMR-PRE experiments were performed by Toyama et al. in order to measure the residue specific NS-ESP of the Caprin1 chains at different ATP and Caprin1 concentrations \cite{Toyama2022}. More specifically, in the absence of ATP molecules, Toyama and co-workers observed a positive ESP with higher values near the regions that contain multiple arginine residues. As the ATP concentration increases to $\sim$1 mM, a pronounced reduction of the NS-ESP of Caprin1 was detected, although the proteins remain in a well-mixed state \cite{Toyama2022}. Over a similar concentration range of ATP the Martini CG model finds that Caprin1 is in a mixed state, and thus we further validate our simulations by calculating the NS-ESP for each $^{1}$H nucleus of each Caprin1 backbone residue. The NS-ESP sums over the contributions of the partial charges from all $\sim$5 million atoms including the other Caprin1 proteins, sodium counterions, and ATP molecules using the following equation: 

\begin{equation}
\phi_{ENS} = - \dfrac{k_{B} T}{2e} ln\left(exp [-\frac{\Delta\ U_{ENS}}{k_{B} T}] \right) 
\end{equation}

\noindent
where $\phi_{ENS}$ is the per-residue NS-ESP, $\Delta$U$_{ENS}$ is the difference in potential energy between the positive and negative probe interactions with Caprin1, $k_{B}$ is the Boltzmann constant, T is the temperature and e is the elementary charge \cite{Yu2021}. Further details are described in Methods. 

\begin{figure}[h]
\centering
\includegraphics[width=0.99\textwidth]{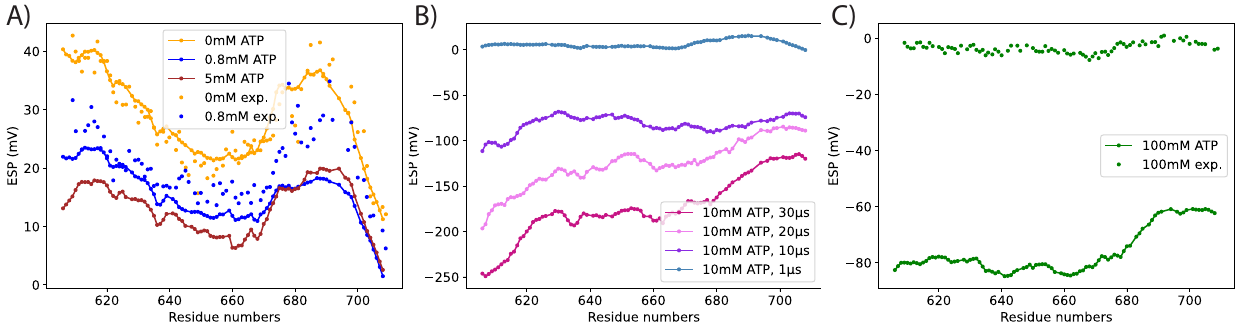}
\caption{\textit{NS-ESP calculations for Caprin1-ATP systems across increasing ATP concentrations.} NS-ESP graphs for Caprin1 \textbf{(A)} in the early mixed state system with 0, 0.8 mM, and 5 mM ATP concentrations compared with experiments of Toyama et al.\cite{Toyama2022} \textbf{B)} the condensate system at 10 mM ATP, here shown as a temporal evolution of the simulations at 1 to 30 $\mu$s, and \textbf{C)} disrupting the condensate with 100 mM ATP concentration in black with experimental values in green. To ensure consistency with the experimental conditions, the charge of the Mg$^{+2}$-free ATP molecules was set to ${-}$3e as stated by Toyama and co-workers \cite{Toyama2022} and the NS-ESP values were scaled down by a factor of 1.8 to account for the experimental buffer. Multiple snapshots were taken every 10 ns at the end of the 30 $\mu$s simulations for all the systems and the NS-ESP calculations were averaged over those snapshots and over all Caprin1 chains.}
\label{fig:esp_plots}
\end{figure}

The NS-ESP of the Caprin1 chains in the initial mixed state is depicted in Figure \ref{fig:esp_plots}A, showing positive NS-ESP values of 10-40 mV depending on ATP concentration and position along the Caprin1 sequence, results that are in excellent agreement with the experimental values \cite{Toyama2022}. At the moderate ATP concentration of 10 mM in which the nanodroplet is formed, we performed a temporal analysis of the NS-ESP evolution (at 1, 10, 20 and 30 $\mu$s simulation time) starting with all Caprin1 chains in the mixed state and gradually capturing the formation of the condensate, as shown in Figure \ref{fig:esp_plots}B. When all Caprin1 chains are soluble, the computed NS-ESP of Caprin1 exhibits positive values, however it gradually transitions to large negative values of several hundred mVs. Notably, regions proximal to the N-terminal region of the Caprin1 chains exhibit larger negative magnitudes compared to the C-terminal region in line with its preference for interactions with ATP molecules. The dissolution of the condensate, invoked by increasing the ATP concentration to 100 mM, systematically reduces the NS-ESP to smaller but still negative values as seen in Figure \ref{fig:esp_plots}C, since both the Caprin1 N-terminus and C-terminus are soluble in water and thus screened to reduced the strong electrostatic interactions present in the condensate.

\begin{figure}[h!]
\centering
\includegraphics[width=0.99\textwidth]{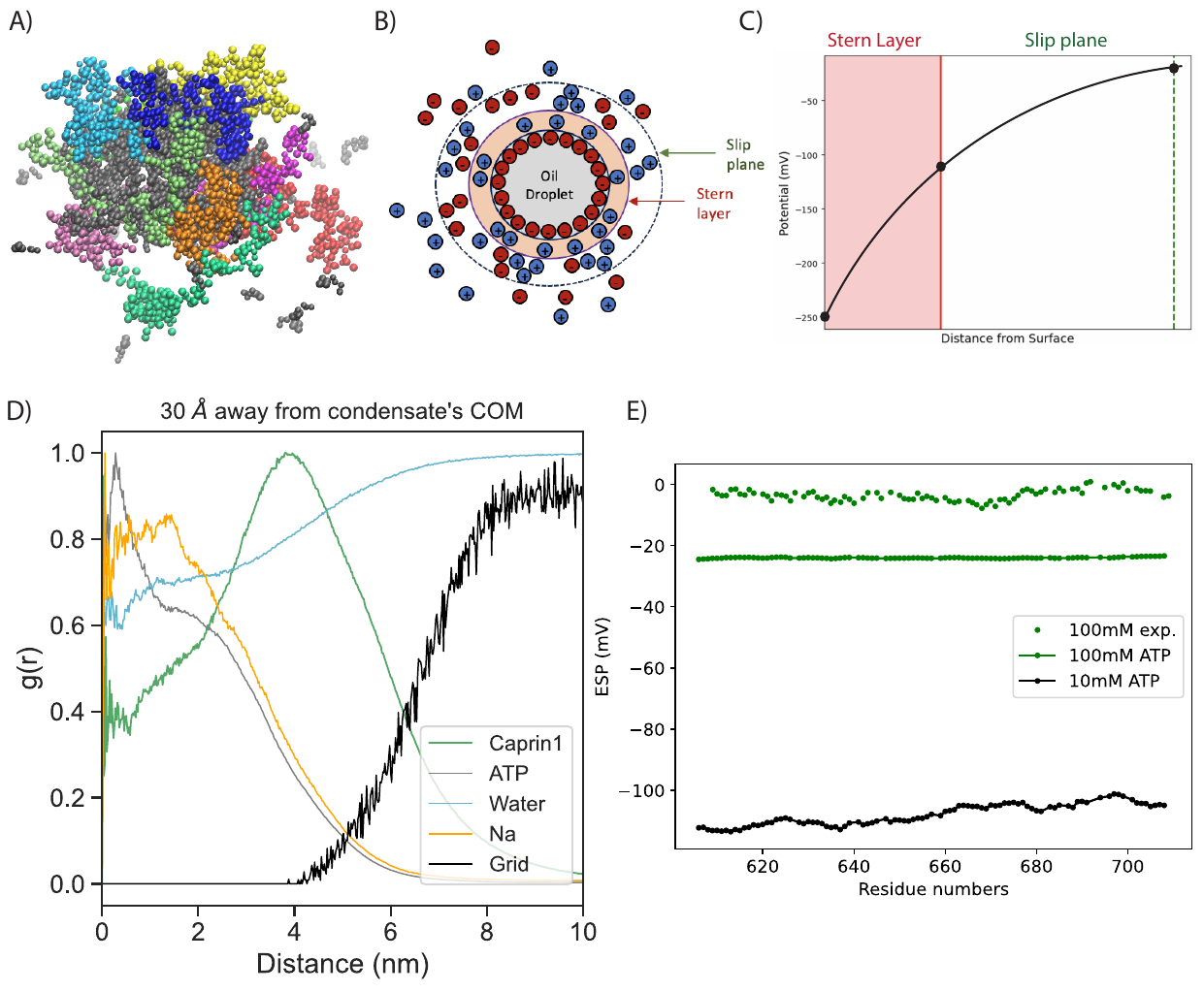}
\caption{\textit{Estimates of the zeta potential for Caprin1-ATP systems for making and breaking the condensate.} \textbf{(A)} the condensate organizes its charge in a way that bears strong similarity to \textbf{(B)} an oil-in-water emulsion. \textbf{(C)} The NS-ESP we calculate for the condensate is an estimate of the Stern potential corresponding to surface ATP and absorbed N-terminus region of the Caprin1-protein; the C-terminus of the Caprin1 chain resides in between the Stern layer and the slip plane which defines the zeta potential. \textbf{D)} the density of charge of all system components including ATP, Na ions, water, and Caprin1 protein, and  placement of grid points at a distance commensurate with the slip plane in B). \textbf{E)} Estimates of the zeta potential for the condensate at 10mM ATP (black); after its dissolution with 100 mM ATP concentration (green), the latter compared with experiment (green dots). Figures \textbf{(B,C)} reproduced from [\cite{LaCour2025}]. Copyright 2025 American Chemical Society. }
\label{fig:zeta_plots}
\end{figure}

The NMR-PRE experiments performed by Toyama et al. for the condensate at moderate levels of ATP were collected after fusion of many microscopic droplets, in which coalescence was driven by centrifugation \cite{Toyama2022}, and are not of the nanocondensate size scale. While these experimental studies help quantify the NS-ESP around Caprin1 in a concentrated and homogeneous phase, they do not directly address the heterogeneous environments present at the surface that stabilizes the original nanodroplets formed in early condensation. In fact it has been proposed that the phase-separated liquid droplets of a condensate carry a surface charge that helps it remain stable against coalescence, analogous to oil-water emulsions \cite{Welsh2022}. We believe that the analogy to oil-water emulsions is worth considering in order to undersrand stability of the simulated Caprin1-ATP nanocondensate in both its formation and during its dissolution. 

The large NS-ESP that we calculate for the Caprin1-ATP condensate (Figure \ref{fig:zeta_plots}A) is equivalent to a charge double layer, or Stern layer, as seen in Figure \ref{fig:zeta_plots}B for an oil-in-water emulsion. For the Caprin1-ATP condensate the Stern layer is defined by an inner layer of negative ATP charge and surrounded by positive charge arising from the N-terminus of the Caprin1 chain. The theory of electrophoresis \cite{anderson1989colloid} states that outside the Stern layer are additional ions that also remain immobilized within the "slip plane", which for our system would correspond to the additional charges of the C-terminus region of Caprin1. Right outside these defined immobilized charge layers for the Caprin1-ATP system is an electric potential that defines the so-called zeta potential. The zeta potential can be measured via its relationship to droplet mobility when an electric field is applied, experiments that have been recently reported for FUS and PR25:PolyU biological condensates \cite{Welsh2022}, and corresponds to an exponential decay of the ESP relative to the NS-ESP measured along the Caprin1 chain as shown in Figure \ref{fig:zeta_plots}C. We thus provide a theoretical estimate of the zeta potential for the Caprin1-ATP condensate, before and after its dissolution, by defining grid points for the ESP calculation outside the region of immobilized charge, i.e. 30-40 $\AA$ from the condensate's center of mass as supported by Figure \ref{fig:zeta_plots}D and similar analysis is performed for the broken condensate as provided in Supplementary Fig. 2. This allows us to estimate the zeta potential for both the formed and broken condensate as seen in Figure \ref{fig:zeta_plots}E. We find that the zeta potential is $\sim$-110 mV for the condensate at 10 mM ATP, a value consistent with it being a stable droplet resistant to droplet coalescence, while increases in ATP from 10 mM to 100 mM lowers the zeta potential to $\sim$-25 mV, consistent with loss of droplet stability that encourages mixing or fusion, and reentry to the mixed state.

\section*{Discussion and Conclusion}
Many early studies of biological condensates at the micron scale have organizations of phase-separated proteins that retain a liquid-like behavior with fast exchange\cite{Liyanage2024}. However there is growing evidence that in cellular environments biomolecular condensates form and are stabilized at much smaller lengthscales of tens to hundreds of nanometers, well below the diffraction limit of standard imaging technologies\cite{Keber2024}. Many biological nanocondensates at these smaller lengthscales are stabilized by co-solutes such as RNA and can exhibit reduced fluidity of a solid- or gel-like assembly\cite{Liyanage2024}. 

Our multiscale simulation methodology is able to reach the same lengthscales of more recent supramolecular resolution experiments, since our CG simulated Caprin1-ATP droplet is $\sim$15-20 nm in diameter, but which allows us to comment on several related observations seen experimentally but with even greater (i.e. atomic) resolution through back-mapping. We find that ATP can act not only as a hydrotrope for dissolving condensates\cite{Patel2017}, but like RNA it can play a role in stabilizing small condensates at appropriate concentrations. Furthermore the formation of ATP clusters within the condensate, stabilized by pi-pi interactions and sodium counterions, are solid-like in composition. Small condensates, at least as far as we observe for the Caprin1-ATP system, are stabilized by the positively charged Caprin1 proteins that are absorbed onto the surface of the ATP core, suggesting an organization of protein in phase separation at these smaller lengthscales to be surface dominated. The structural organization of the Caprin1-ATP nanodroplet is thus unusual compared to other reported assemblies of biological condensates at larger lengthscales, in which proteins associate internally within the condensate envelope\cite{Banani2017}. 

Furthermore we believe this study also addresses the stability of these small condensates as aided by ATP. It seems to us that analogies to oil-in-water emulsions and the role of electrostatics is a fruitful line of investigation to address stability. Hence our evaluation of the NS-ESP for the early mixed state, in excellent agreement with NMR-PRE experiments, gave us an important validation of the accuracy of our simulated result, and thus permitting us to predict the NS-ESP of the small and spontaneously formed condensate at moderate ATP concentrations and its dissolution at much higher ATP concentrations. Given the structural organization of the nanoscale condensate we observe in our simulations, the NS-ESP is a good measure of a Stern potential involving a double layer of a negative ATP core interacting with the positively charged N-terminus of the Caprin1 protein. Additional charges from the C-terminus of Caprin1 are in the slip plane, and are defined to be "immobile" in the sense that they are connected through covalency of the protein chain. I.e. under an applied electric field the whole Caprin1-ATP assembly would move toward (likely) the anode as measured electrokinetically. 

By placing grid points outside the region of immobile charge, our evaluation of the ESP can be taken as an estimate of the zeta potential further from the surfaces of the Caprin1-ATP nanodroplet condensate, and as it dissolves at higher ATP concentrations. This yields an estimate of $\sim$-110 mV for the zeta potential for the small Caprin1-ATP condensate, a value commensurate with a stable oil emulsion.\cite{Welsh2022} As ATP concentration increases and the structural organization changes due to weakened interactions between ATP and Caprin1, the zeta potential reduces to $\sim$-25 mV , and consistent with values that induce flocculation in oil-water mixtures. This is supported by Welsh et al. in which they measured the zeta potential for large condensates, showing that higher condensate surface charge correlated with lower fusion propensity\cite{Welsh2022}. It is interesting to note that the small condensates in our simulations have larger zeta potentials compared to the zeta potentials experimentally measured by Welsh and co-workers for micron scale sized condensates of FUS and PR25:PolyU\cite{Welsh2022}. This may be due to limitations of our force field model of course, but it nonetheless calls for experimental measurement of zeta potentials for small condensates to test this possibility and to understand their relative stability across the whole size scale distribution of biomolecular condensates. 

\subsection*{Supporting Information Appendix (SI)}
Figures S1, S2 and S3 show details of ATP oligomerization and probe accessibility into the condensate.

\section{Acknowledgments}
{We gratefully acknowledge funding from the National Institutes of Health (2R01GM127627-05). We thank Julie Forman-Kay for may interesting discussions. M. Tsanai thanks Dr. Selim Sami for help with the ESP calculations. In honor of Charles Brooks III and his many seminal contributions to the field of biomolecular simulation.}

\bibliography{references}

\providecommand{\latin}[1]{#1}
\makeatletter
\providecommand{\doi}
  {\begingroup\let\do\@makeother\dospecials
  \catcode`\{=1 \catcode`\}=2 \doi@aux}
\providecommand{\doi@aux}[1]{\endgroup\texttt{#1}}
\makeatother
\providecommand*\mcitethebibliography{\thebibliography}
\csname @ifundefined\endcsname{endmcitethebibliography}  {\let\endmcitethebibliography\endthebibliography}{}
\begin{mcitethebibliography}{77}
\providecommand*\natexlab[1]{#1}
\providecommand*\mciteSetBstSublistMode[1]{}
\providecommand*\mciteSetBstMaxWidthForm[2]{}
\providecommand*\mciteBstWouldAddEndPuncttrue
  {\def\EndOfBibitem{\unskip.}}
\providecommand*\mciteBstWouldAddEndPunctfalse
  {\let\EndOfBibitem\relax}
\providecommand*\mciteSetBstMidEndSepPunct[3]{}
\providecommand*\mciteSetBstSublistLabelBeginEnd[3]{}
\providecommand*\EndOfBibitem{}
\mciteSetBstSublistMode{f}
\mciteSetBstMaxWidthForm{subitem}{(\alph{mcitesubitemcount})}
\mciteSetBstSublistLabelBeginEnd
  {\mcitemaxwidthsubitemform\space}
  {\relax}
  {\relax}

\bibitem[Chu \latin{et~al.}(2022)Chu, Xu, Tong, Wang, and Zhang]{Chu2022}
Chu,~X.-Y.; Xu,~Y.-Y.; Tong,~X.-Y.; Wang,~G.; Zhang,~H.-Y. The Legend of ATP: From Origin of Life to Precision Medicine. \emph{Metabolites} \textbf{2022}, \emph{12}, 461\relax
\mciteBstWouldAddEndPuncttrue
\mciteSetBstMidEndSepPunct{\mcitedefaultmidpunct}
{\mcitedefaultendpunct}{\mcitedefaultseppunct}\relax
\EndOfBibitem
\bibitem[Ren \latin{et~al.}(2022)Ren, Shan, Zhang, Ding, and Ma]{Ren2022}
Ren,~C.-L.; Shan,~Y.; Zhang,~P.; Ding,~H.-M.; Ma,~Y.-Q. Uncovering the molecular mechanism for dual effect of ATP on phase separation in FUS solution. \emph{Science Advances} \textbf{2022}, \emph{8}, eabo7885\relax
\mciteBstWouldAddEndPuncttrue
\mciteSetBstMidEndSepPunct{\mcitedefaultmidpunct}
{\mcitedefaultendpunct}{\mcitedefaultseppunct}\relax
\EndOfBibitem
\bibitem[Aida \latin{et~al.}(2022)Aida, Shigeta, and Harada]{Aida2022}
Aida,~H.; Shigeta,~Y.; Harada,~R. The role of ATP in solubilizing RNA-binding protein fused in sarcoma. \emph{Proteins: Structure, Function, and Bioinformatics} \textbf{2022}, \emph{90}, 1606--1612\relax
\mciteBstWouldAddEndPuncttrue
\mciteSetBstMidEndSepPunct{\mcitedefaultmidpunct}
{\mcitedefaultendpunct}{\mcitedefaultseppunct}\relax
\EndOfBibitem
\bibitem[Mehringer \latin{et~al.}(2021)Mehringer, Do, Touraud, Hohenschutz, Khoshsima, Horinek, and Kunz]{Mehringer2021}
Mehringer,~J.; Do,~T.-M.; Touraud,~D.; Hohenschutz,~M.; Khoshsima,~A.; Horinek,~D.; Kunz,~W. Hofmeister versus Neuberg: is ATP really a biological hydrotrope? \emph{Cell Reports Physical Science} \textbf{2021}, \emph{2}, 100343\relax
\mciteBstWouldAddEndPuncttrue
\mciteSetBstMidEndSepPunct{\mcitedefaultmidpunct}
{\mcitedefaultendpunct}{\mcitedefaultseppunct}\relax
\EndOfBibitem
\bibitem[Song(2021)]{Song2021}
Song,~J. Adenosine triphosphate energy-independently controls protein homeostasis with unique structure and diverse mechanisms. \emph{Protein Science} \textbf{2021}, \emph{30}, 1277--1293\relax
\mciteBstWouldAddEndPuncttrue
\mciteSetBstMidEndSepPunct{\mcitedefaultmidpunct}
{\mcitedefaultendpunct}{\mcitedefaultseppunct}\relax
\EndOfBibitem
\bibitem[Traut(1994)]{Traut1994}
Traut,~T.~W. Physiological concentrations of purines and pyrimidines. \emph{Molecular and Cellular Biochemistry} \textbf{1994}, \emph{140}, 1--22\relax
\mciteBstWouldAddEndPuncttrue
\mciteSetBstMidEndSepPunct{\mcitedefaultmidpunct}
{\mcitedefaultendpunct}{\mcitedefaultseppunct}\relax
\EndOfBibitem
\bibitem[Patel \latin{et~al.}(2017)Patel, Malinovska, Saha, Wang, Alberti, Krishnan, and Hyman]{Patel2017}
Patel,~A.; Malinovska,~L.; Saha,~S.; Wang,~J.; Alberti,~S.; Krishnan,~Y.; Hyman,~A.~A. ATP as a biological hydrotrope. \emph{Science} \textbf{2017}, \emph{356}, 753--756\relax
\mciteBstWouldAddEndPuncttrue
\mciteSetBstMidEndSepPunct{\mcitedefaultmidpunct}
{\mcitedefaultendpunct}{\mcitedefaultseppunct}\relax
\EndOfBibitem
\bibitem[Sridharan \latin{et~al.}(2019)Sridharan, Kurzawa, Werner, Günthner, Helm, Huber, Bantscheff, and Savitski]{Sridharan2019}
Sridharan,~S.; Kurzawa,~N.; Werner,~T.; Günthner,~I.; Helm,~D.; Huber,~W.; Bantscheff,~M.; Savitski,~M.~M. Proteome-wide solubility and thermal stability profiling reveals distinct regulatory roles for ATP. \emph{Nature Communications} \textbf{2019}, \emph{10}, 1155\relax
\mciteBstWouldAddEndPuncttrue
\mciteSetBstMidEndSepPunct{\mcitedefaultmidpunct}
{\mcitedefaultendpunct}{\mcitedefaultseppunct}\relax
\EndOfBibitem
\bibitem[Eastoe \latin{et~al.}(2011)Eastoe, Hatzopoulos, and Dowding]{Eastoe2011}
Eastoe,~J.; Hatzopoulos,~M.~H.; Dowding,~P.~J. Action of hydrotropes and alkyl-hydrotropes. \emph{Soft Matter} \textbf{2011}, \emph{7}, 5917--5925\relax
\mciteBstWouldAddEndPuncttrue
\mciteSetBstMidEndSepPunct{\mcitedefaultmidpunct}
{\mcitedefaultendpunct}{\mcitedefaultseppunct}\relax
\EndOfBibitem
\bibitem[Kang \latin{et~al.}(2018)Kang, Lim, and Song]{Kang2018}
Kang,~J.; Lim,~L.; Song,~J. ATP enhances at low concentrations but dissolves at high concentrations liquid-liquid phase separation (LLPS) of ALS/FTD-causing FUS. \emph{Biochemical and Biophysical Research Communications} \textbf{2018}, \emph{504}, 545--551\relax
\mciteBstWouldAddEndPuncttrue
\mciteSetBstMidEndSepPunct{\mcitedefaultmidpunct}
{\mcitedefaultendpunct}{\mcitedefaultseppunct}\relax
\EndOfBibitem
\bibitem[Hu \latin{et~al.}(2022)Hu, Ou, and Li]{Hu2022}
Hu,~G.; Ou,~X.; Li,~J. Mechanistic Insight on General Protein-Binding Ability of ATP and the Impacts of Arginine Residues. \emph{J. Phys. Chem. B} \textbf{2022}, \emph{126}, 4647--4658\relax
\mciteBstWouldAddEndPuncttrue
\mciteSetBstMidEndSepPunct{\mcitedefaultmidpunct}
{\mcitedefaultendpunct}{\mcitedefaultseppunct}\relax
\EndOfBibitem
\bibitem[Zalar \latin{et~al.}(2023)Zalar, Bye, and Curtis]{Zalar2023}
Zalar,~M.; Bye,~J.; Curtis,~R. Nonspecific Binding of Adenosine Tripolyphosphate and Tripolyphosphate Modulates the Phase Behavior of Lysozyme. \emph{J. Am. Chem. Soc.} \textbf{2023}, \emph{145}, 929--943\relax
\mciteBstWouldAddEndPuncttrue
\mciteSetBstMidEndSepPunct{\mcitedefaultmidpunct}
{\mcitedefaultendpunct}{\mcitedefaultseppunct}\relax
\EndOfBibitem
\bibitem[Kota \latin{et~al.}(2024)Kota, Prasad, and Zhou]{Kota2024}
Kota,~D.; Prasad,~R.; Zhou,~H.-X. Adenosine Triphosphate Mediates Phase Separation of Disordered Basic Proteins by Bridging Intermolecular Interaction Networks. \emph{J. Am. Chem. Soc.} \textbf{2024}, \emph{146}, 1326--1336\relax
\mciteBstWouldAddEndPuncttrue
\mciteSetBstMidEndSepPunct{\mcitedefaultmidpunct}
{\mcitedefaultendpunct}{\mcitedefaultseppunct}\relax
\EndOfBibitem
\bibitem[Liu and Wang(2023)Liu, and Wang]{Liu2023}
Liu,~F.; Wang,~J. ATP Acts as a Hydrotrope to Regulate the Phase Separation of NBDY Clusters. \emph{JACS Au} \textbf{2023}, \emph{3}, 2578--2585\relax
\mciteBstWouldAddEndPuncttrue
\mciteSetBstMidEndSepPunct{\mcitedefaultmidpunct}
{\mcitedefaultendpunct}{\mcitedefaultseppunct}\relax
\EndOfBibitem
\bibitem[Wright and Dyson(2015)Wright, and Dyson]{Wright2015}
Wright,~P.~E.; Dyson,~H.~J. Intrinsically disordered proteins in cellular signalling and regulation. \emph{Nature Reviews Molecular Cell Biology} \textbf{2015}, \emph{16}, 18--29\relax
\mciteBstWouldAddEndPuncttrue
\mciteSetBstMidEndSepPunct{\mcitedefaultmidpunct}
{\mcitedefaultendpunct}{\mcitedefaultseppunct}\relax
\EndOfBibitem
\bibitem[Boeynaems \latin{et~al.}(2018)Boeynaems, Alberti, Fawzi, Mittag, Polymenidou, Rousseau, Schymkowitz, Shorter, Wolozin, {Van Den Bosch}, Tompa, and Fuxreiter]{Boeynaems2018}
Boeynaems,~S.; Alberti,~S.; Fawzi,~N.~L.; Mittag,~T.; Polymenidou,~M.; Rousseau,~F.; Schymkowitz,~J.; Shorter,~J.; Wolozin,~B.; {Van Den Bosch},~L.; Tompa,~P.; Fuxreiter,~M. Protein Phase Separation: A New Phase in Cell Biology. \emph{Trends in Cell Biology} \textbf{2018}, \emph{28}, 420--435\relax
\mciteBstWouldAddEndPuncttrue
\mciteSetBstMidEndSepPunct{\mcitedefaultmidpunct}
{\mcitedefaultendpunct}{\mcitedefaultseppunct}\relax
\EndOfBibitem
\bibitem[Brady \latin{et~al.}(2017)Brady, Farber, Sekhar, Lin, Huang, Bah, Nott, Chan, Baldwin, Forman-Kay, and Kay]{Brady2017}
Brady,~J.~P.; Farber,~P.~J.; Sekhar,~A.; Lin,~Y.-H.; Huang,~R.; Bah,~A.; Nott,~T.~J.; Chan,~H.~S.; Baldwin,~A.~J.; Forman-Kay,~J.~D.; Kay,~L.~E. Structural and hydrodynamic properties of an intrinsically disordered region of a germ cell-specific protein on phase separation. \emph{Proceedings of the National Academy of Sciences} \textbf{2017}, \emph{114}, E8194--E8203\relax
\mciteBstWouldAddEndPuncttrue
\mciteSetBstMidEndSepPunct{\mcitedefaultmidpunct}
{\mcitedefaultendpunct}{\mcitedefaultseppunct}\relax
\EndOfBibitem
\bibitem[Qamar \latin{et~al.}(2018)Qamar, Wang, Randle, Ruggeri, Varela, Lin, Phillips, Miyashita, Williams, Str$\''o$hl, Meadows, Ferry, Dardov, Tartaglia, Farrer, Kaminski~Schierle, Kaminski, Holt, Fraser, Schmitt-Ulms, Klenerman, nowles, Vendruscolo, and St~George-Hyslop]{Qamar2018}
Qamar,~S. \latin{et~al.}  FUS Phase Separation Is Modulated by a Molecular Chaperone and Methylation of Arginine Cation-$\pi$; Interactions. \emph{Cell} \textbf{2018}, \emph{173}, 720--734.e15\relax
\mciteBstWouldAddEndPuncttrue
\mciteSetBstMidEndSepPunct{\mcitedefaultmidpunct}
{\mcitedefaultendpunct}{\mcitedefaultseppunct}\relax
\EndOfBibitem
\bibitem[Wang \latin{et~al.}(2018)Wang, Choi, Holehouse, Lee, Zhang, Jahnel, Maharana, Lemaitre, Pozniakovsky, Drechsel, Poser, Pappu, Alberti, and Hyman]{Wang2018}
Wang,~J.; Choi,~J.-M.; Holehouse,~A.~S.; Lee,~H.~O.; Zhang,~X.; Jahnel,~M.; Maharana,~S.; Lemaitre,~R.; Pozniakovsky,~A.; Drechsel,~D.; Poser,~I.; Pappu,~R.~V.; Alberti,~S.; Hyman,~A.~A. A Molecular Grammar Governing the Driving Forces for Phase Separation of Prion-like RNA Binding Proteins. \emph{Cell} \textbf{2018}, \emph{174}, 688--699.e16\relax
\mciteBstWouldAddEndPuncttrue
\mciteSetBstMidEndSepPunct{\mcitedefaultmidpunct}
{\mcitedefaultendpunct}{\mcitedefaultseppunct}\relax
\EndOfBibitem
\bibitem[Vernon \latin{et~al.}(2018)Vernon, Chong, Tsang, Kim, Bah, Farber, Lin, and Forman-Kay]{Vernon2018}
Vernon,~R.~M.; Chong,~P.~A.; Tsang,~B.; Kim,~T.~H.; Bah,~A.; Farber,~P.; Lin,~H.; Forman-Kay,~J.~D. Pi-Pi contacts are an overlooked protein feature relevant to phase separation. \emph{eLife} \textbf{2018}, \emph{7}, e31486\relax
\mciteBstWouldAddEndPuncttrue
\mciteSetBstMidEndSepPunct{\mcitedefaultmidpunct}
{\mcitedefaultendpunct}{\mcitedefaultseppunct}\relax
\EndOfBibitem
\bibitem[Carter-Fenk \latin{et~al.}(2023)Carter-Fenk, Liu, Pujal, Loipersberger, Tsanai, Vernon, Forman-Kay, Head-Gordon, Heidar-Zadeh, and Head-Gordon]{Carter-Fenk2023}
Carter-Fenk,~K.; Liu,~M.; Pujal,~L.; Loipersberger,~M.; Tsanai,~M.; Vernon,~R.~M.; Forman-Kay,~J.~D.; Head-Gordon,~M.; Heidar-Zadeh,~F.; Head-Gordon,~T. The Energetic Origins of Pi–Pi Contacts in Proteins. \emph{Journal of the American Chemical Society} \textbf{2023}, \emph{145}, 24836--24851\relax
\mciteBstWouldAddEndPuncttrue
\mciteSetBstMidEndSepPunct{\mcitedefaultmidpunct}
{\mcitedefaultendpunct}{\mcitedefaultseppunct}\relax
\EndOfBibitem
\bibitem[Murthy \latin{et~al.}(2019)Murthy, Dignon, Kan, Zerze, Parekh, Mittal, and Fawzi]{Murthy2019}
Murthy,~A.~C.; Dignon,~G.~L.; Kan,~Y.; Zerze,~G.~H.; Parekh,~S.~H.; Mittal,~J.; Fawzi,~N.~L. Molecular interactions underlying liquid-liquid phase separation of the FUS low-complexity domain. \emph{Nature Structural \& Molecular Biology} \textbf{2019}, \emph{26}, 637--648\relax
\mciteBstWouldAddEndPuncttrue
\mciteSetBstMidEndSepPunct{\mcitedefaultmidpunct}
{\mcitedefaultendpunct}{\mcitedefaultseppunct}\relax
\EndOfBibitem
\bibitem[Wilson and Chin(1991)Wilson, and Chin]{Wilson1991}
Wilson,~J.~E.; Chin,~A. Chelation of divalent cations by ATP, studied by titration calorimetry. \emph{Analytical Biochemistry} \textbf{1991}, \emph{193}, 16--19\relax
\mciteBstWouldAddEndPuncttrue
\mciteSetBstMidEndSepPunct{\mcitedefaultmidpunct}
{\mcitedefaultendpunct}{\mcitedefaultseppunct}\relax
\EndOfBibitem
\bibitem[Ou \latin{et~al.}(2021)Ou, Lao, Xu, Wutthinitikornkit, Shi, Chen, and Li]{Ou2021}
Ou,~X.; Lao,~Y.; Xu,~J.; Wutthinitikornkit,~Y.; Shi,~R.; Chen,~X.; Li,~J. ATP Can Efficiently Stabilize Protein through a Unique Mechanism. \emph{JACS Au} \textbf{2021}, \emph{1}, 1766--1777\relax
\mciteBstWouldAddEndPuncttrue
\mciteSetBstMidEndSepPunct{\mcitedefaultmidpunct}
{\mcitedefaultendpunct}{\mcitedefaultseppunct}\relax
\EndOfBibitem
\bibitem[Nishizawa \latin{et~al.}(2021)Nishizawa, Walinda, Morimoto, Kohn, Scheler, Shirakawa, and Sugase]{Nishizawa2021}
Nishizawa,~M.; Walinda,~E.; Morimoto,~D.; Kohn,~B.; Scheler,~U.; Shirakawa,~M.; Sugase,~K. Effects of Weak Nonspecific Interactions with ATP on Proteins. \emph{J. Am. Chem. Soc.} \textbf{2021}, \emph{143}, 11982--11993\relax
\mciteBstWouldAddEndPuncttrue
\mciteSetBstMidEndSepPunct{\mcitedefaultmidpunct}
{\mcitedefaultendpunct}{\mcitedefaultseppunct}\relax
\EndOfBibitem
\bibitem[Choi \latin{et~al.}(2024)Choi, Zhou, Tabo, Heald, and Xu]{Choi2024}
Choi,~A.~A.; Zhou,~C.~Y.; Tabo,~A.; Heald,~R.; Xu,~K. Single-molecule diffusivity quantification in Xenopus egg extracts elucidates physicochemical properties of the cytoplasm. \emph{Proc Natl Acad Sci U S A} \textbf{2024}, \emph{121}, e2411402121\relax
\mciteBstWouldAddEndPuncttrue
\mciteSetBstMidEndSepPunct{\mcitedefaultmidpunct}
{\mcitedefaultendpunct}{\mcitedefaultseppunct}\relax
\EndOfBibitem
\bibitem[Sigel and Griesser(2005)Sigel, and Griesser]{Sigel2005}
Sigel,~H.; Griesser,~R. Nucleoside 5-triphosphates: self-association, acid–base, and metal ion-binding properties in solution. \emph{Chem. Soc. Rev.} \textbf{2005}, \emph{34}, 875--900\relax
\mciteBstWouldAddEndPuncttrue
\mciteSetBstMidEndSepPunct{\mcitedefaultmidpunct}
{\mcitedefaultendpunct}{\mcitedefaultseppunct}\relax
\EndOfBibitem
\bibitem[Banani \latin{et~al.}(2017)Banani, Lee, Hyman, and Rosen]{Banani2017}
Banani,~S.~F.; Lee,~H.~O.; Hyman,~A.~A.; Rosen,~M.~K. Biomolecular condensates: organizers of cellular biochemistry. \emph{Nature Reviews Molecular Cell Biology} \textbf{2017}, \emph{18}, 285--295\relax
\mciteBstWouldAddEndPuncttrue
\mciteSetBstMidEndSepPunct{\mcitedefaultmidpunct}
{\mcitedefaultendpunct}{\mcitedefaultseppunct}\relax
\EndOfBibitem
\bibitem[Gomes and Shorter(2019)Gomes, and Shorter]{Gomes2019}
Gomes,~E.; Shorter,~J. The molecular language of membraneless organelles. \emph{Journal of Biological Chemistry} \textbf{2019}, \emph{294}, 7115--7127\relax
\mciteBstWouldAddEndPuncttrue
\mciteSetBstMidEndSepPunct{\mcitedefaultmidpunct}
{\mcitedefaultendpunct}{\mcitedefaultseppunct}\relax
\EndOfBibitem
\bibitem[Tsanai \latin{et~al.}(2021)Tsanai, Frederix, Schroer, Souza, and Marrink]{Tsanai2021}
Tsanai,~M.; Frederix,~P. W. J.~M.; Schroer,~C. F.~E.; Souza,~P. C.~T.; Marrink,~S.~J. Coacervate formation studied by explicit solvent coarse-grain molecular dynamics with the Martini model. \emph{Chem. Sci.} \textbf{2021}, \emph{12}, 8521--8530\relax
\mciteBstWouldAddEndPuncttrue
\mciteSetBstMidEndSepPunct{\mcitedefaultmidpunct}
{\mcitedefaultendpunct}{\mcitedefaultseppunct}\relax
\EndOfBibitem
\bibitem[Brasnett \latin{et~al.}(2024)Brasnett, Kiani, Sami, Otto, and Marrink]{Brasnett2024}
Brasnett,~C.; Kiani,~A.; Sami,~S.; Otto,~S.; Marrink,~S.~J. Capturing chemical reactions inside biomolecular condensates with reactive Martini simulations. \emph{Communications Chemistry} \textbf{2024}, \emph{7}, 151\relax
\mciteBstWouldAddEndPuncttrue
\mciteSetBstMidEndSepPunct{\mcitedefaultmidpunct}
{\mcitedefaultendpunct}{\mcitedefaultseppunct}\relax
\EndOfBibitem
\bibitem[Posey \latin{et~al.}(2024)Posey, Bremer, Erkamp, Pant, Knowles, Dai, Mittag, and Pappu]{Posey2024}
Posey,~A.~E.; Bremer,~A.; Erkamp,~N.~A.; Pant,~A.; Knowles,~T. P.~J.; Dai,~Y.; Mittag,~T.; Pappu,~R.~V. Biomolecular Condensates are Characterized by Interphase Electric Potentials. \emph{Journal of the American Chemical Society} \textbf{2024}, \emph{146}, 28268--28281\relax
\mciteBstWouldAddEndPuncttrue
\mciteSetBstMidEndSepPunct{\mcitedefaultmidpunct}
{\mcitedefaultendpunct}{\mcitedefaultseppunct}\relax
\EndOfBibitem
\bibitem[Yu \latin{et~al.}(2021)Yu, Pletka, Pettitt, and Iwahara]{Yu2021}
Yu,~B.; Pletka,~C.~C.; Pettitt,~B.~M.; Iwahara,~J. De novo determination of near-surface electrostatic potentials by NMR. \emph{Proceedings of the National Academy of Sciences} \textbf{2021}, \emph{118}, e2104020118\relax
\mciteBstWouldAddEndPuncttrue
\mciteSetBstMidEndSepPunct{\mcitedefaultmidpunct}
{\mcitedefaultendpunct}{\mcitedefaultseppunct}\relax
\EndOfBibitem
\bibitem[Toyama \latin{et~al.}(2022)Toyama, Rangadurai, Forman-Kay, and Kay]{Toyama2022}
Toyama,~Y.; Rangadurai,~A.~K.; Forman-Kay,~J.~D.; Kay,~L.~E. Mapping the per-residue surface electrostatic potential of CAPRIN1 along its phase-separation trajectory. \emph{Proceedings of the National Academy of Sciences} \textbf{2022}, \emph{119}, e2210492119\relax
\mciteBstWouldAddEndPuncttrue
\mciteSetBstMidEndSepPunct{\mcitedefaultmidpunct}
{\mcitedefaultendpunct}{\mcitedefaultseppunct}\relax
\EndOfBibitem
\bibitem[Lin \latin{et~al.}(2025)Lin, Kim, Das, Pal, Wessén, Rangadurai, Kay, Forman-Kay, and Chan]{Lin2024}
Lin,~Y.-H.; Kim,~T.~H.; Das,~S.; Pal,~T.; Wessén,~J.; Rangadurai,~A.~K.; Kay,~L.~E.; Forman-Kay,~J.~D.; Chan,~H.~S. Electrostatics of salt-dependent reentrant phase behaviors highlights diverse roles of ATP in biomolecular condensates. \emph{eLife} \textbf{2025}, \emph{13}, RP100284\relax
\mciteBstWouldAddEndPuncttrue
\mciteSetBstMidEndSepPunct{\mcitedefaultmidpunct}
{\mcitedefaultendpunct}{\mcitedefaultseppunct}\relax
\EndOfBibitem
\bibitem[Welsh \latin{et~al.}(2022)Welsh, Krainer, Espinosa, Joseph, Sridhar, Jahnel, Arter, Saar, Alberti, Collepardo-Guevara, and Knowles]{Welsh2022}
Welsh,~T.~J.; Krainer,~G.; Espinosa,~J.~R.; Joseph,~J.~A.; Sridhar,~A.; Jahnel,~M.; Arter,~W.~E.; Saar,~K.~L.; Alberti,~S.; Collepardo-Guevara,~R.; Knowles,~T. P.~J. Surface Electrostatics Govern the Emulsion Stability of Biomolecular Condensates. \emph{Nano Letters} \textbf{2022}, \emph{22}, 612--621\relax
\mciteBstWouldAddEndPuncttrue
\mciteSetBstMidEndSepPunct{\mcitedefaultmidpunct}
{\mcitedefaultendpunct}{\mcitedefaultseppunct}\relax
\EndOfBibitem
\bibitem[Krainer \latin{et~al.}(2021)Krainer, Welsh, Joseph, Espinosa, Wittmann, de~Csilléry, Sridhar, Toprakcioglu, Gudiškytė, Czekalska, Arter, Guillén-Boixet, Franzmann, Qamar, George-Hyslop, Hyman, Collepardo-Guevara, Alberti, and Knowles]{Krainer2021}
Krainer,~G. \latin{et~al.}  Reentrant liquid condensate phase of proteins is stabilized by hydrophobic and non-ionic interactions. \emph{Nature Communications} \textbf{2021}, \emph{12}, 1085\relax
\mciteBstWouldAddEndPuncttrue
\mciteSetBstMidEndSepPunct{\mcitedefaultmidpunct}
{\mcitedefaultendpunct}{\mcitedefaultseppunct}\relax
\EndOfBibitem
\bibitem[Martin \latin{et~al.}(2021)Martin, Harmon, Hopkins, Chakravarthy, Incicco, Schuck, Soranno, and Mittag]{Martin2021}
Martin,~E.~W.; Harmon,~T.~S.; Hopkins,~J.~B.; Chakravarthy,~S.; Incicco,~J.~J.; Schuck,~P.; Soranno,~A.; Mittag,~T. A multi-step nucleation process determines the kinetics of prion-like domain phase separation. \emph{Nature Communications} \textbf{2021}, \emph{12}, 4513\relax
\mciteBstWouldAddEndPuncttrue
\mciteSetBstMidEndSepPunct{\mcitedefaultmidpunct}
{\mcitedefaultendpunct}{\mcitedefaultseppunct}\relax
\EndOfBibitem
\bibitem[Forman-Kay \latin{et~al.}(2022)Forman-Kay, Ditlev, Nosella, and Lee]{JFK2022}
Forman-Kay,~J.~D.; Ditlev,~J.~A.; Nosella,~M.~L.; Lee,~H.~O. What are the distinguishing features and size requirements of biomolecular condensates and their implications for RNA-containing condensates? \emph{Rna} \textbf{2022}, \emph{28}, 36--47\relax
\mciteBstWouldAddEndPuncttrue
\mciteSetBstMidEndSepPunct{\mcitedefaultmidpunct}
{\mcitedefaultendpunct}{\mcitedefaultseppunct}\relax
\EndOfBibitem
\bibitem[Cho \latin{et~al.}(2018)Cho, Spille, Hecht, Lee, Li, Grube, and Cisse]{Cho2018}
Cho,~W.-K.; Spille,~J.-H.; Hecht,~M.; Lee,~C.; Li,~C.; Grube,~V.; Cisse,~I.~I. Mediator and RNA polymerase II clusters associate in transcription-dependent condensates. \emph{Science} \textbf{2018}, \emph{361}, 412--415\relax
\mciteBstWouldAddEndPuncttrue
\mciteSetBstMidEndSepPunct{\mcitedefaultmidpunct}
{\mcitedefaultendpunct}{\mcitedefaultseppunct}\relax
\EndOfBibitem
\bibitem[Rey \latin{et~al.}(2020)Rey, Zaganelli, Cuillery, Vartholomaiou, Croisier, Martinou, and Manley]{Rey2020}
Rey,~T.; Zaganelli,~S.; Cuillery,~E.; Vartholomaiou,~E.; Croisier,~M.; Martinou,~J.-C.; Manley,~S. Mitochondrial RNA granules are fluid condensates positioned by membrane dynamics. \emph{Nature Cell Biology} \textbf{2020}, \emph{22}, 1180--1186\relax
\mciteBstWouldAddEndPuncttrue
\mciteSetBstMidEndSepPunct{\mcitedefaultmidpunct}
{\mcitedefaultendpunct}{\mcitedefaultseppunct}\relax
\EndOfBibitem
\bibitem[Keber \latin{et~al.}(2024)Keber, Nguyen, Mariossi, Brangwynne, and Wühr]{Keber2024}
Keber,~F.~C.; Nguyen,~T.; Mariossi,~A.; Brangwynne,~C.~P.; Wühr,~M. Evidence for widespread cytoplasmic structuring into mesoscale condensates. \emph{Nature Cell Biology} \textbf{2024}, \emph{26}, 346--352\relax
\mciteBstWouldAddEndPuncttrue
\mciteSetBstMidEndSepPunct{\mcitedefaultmidpunct}
{\mcitedefaultendpunct}{\mcitedefaultseppunct}\relax
\EndOfBibitem
\bibitem[Kar \latin{et~al.}(2022)Kar, Dar, Welsh, Vogel, Kühnemuth, Majumdar, Krainer, Franzmann, Alberti, Seidel, Knowles, Hyman, and Pappu]{Kar2022}
Kar,~M.; Dar,~F.; Welsh,~T.~J.; Vogel,~L.~T.; Kühnemuth,~R.; Majumdar,~A.; Krainer,~G.; Franzmann,~T.~M.; Alberti,~S.; Seidel,~C. A.~M.; Knowles,~T. P.~J.; Hyman,~A.~A.; Pappu,~R.~V. Phase-separating RNA-binding proteins form heterogeneous distributions of clusters in subsaturated solutions. \emph{Proceedings of the National Academy of Sciences} \textbf{2022}, \emph{119}, e2202222119\relax
\mciteBstWouldAddEndPuncttrue
\mciteSetBstMidEndSepPunct{\mcitedefaultmidpunct}
{\mcitedefaultendpunct}{\mcitedefaultseppunct}\relax
\EndOfBibitem
\bibitem[Kar \latin{et~al.}(2024)Kar, Vogel, Chauhan, Felekyan, Ausserwöger, Welsh, Dar, Kamath, Knowles, Hyman, Seidel, and Pappu]{Kar2024}
Kar,~M.; Vogel,~L.~T.; Chauhan,~G.; Felekyan,~S.; Ausserwöger,~H.; Welsh,~T.~J.; Dar,~F.; Kamath,~A.~R.; Knowles,~T. P.~J.; Hyman,~A.~A.; Seidel,~C. A.~M.; Pappu,~R.~V. Solutes unmask differences in clustering versus phase separation of FET proteins. \emph{Nature Communications} \textbf{2024}, \emph{15}, 4408\relax
\mciteBstWouldAddEndPuncttrue
\mciteSetBstMidEndSepPunct{\mcitedefaultmidpunct}
{\mcitedefaultendpunct}{\mcitedefaultseppunct}\relax
\EndOfBibitem
\bibitem[Gil-Garcia \latin{et~al.}(2024)Gil-Garcia, Benítez-Mateos, Papp, Stoffel, Morelli, Normak, Makasewicz, Faltova, Paradisi, and Arosio]{GilGarcia2024}
Gil-Garcia,~M.; Benítez-Mateos,~A.~I.; Papp,~M.; Stoffel,~F.; Morelli,~C.; Normak,~K.; Makasewicz,~K.; Faltova,~L.; Paradisi,~F.; Arosio,~P. Local environment in biomolecular condensates modulates enzymatic activity across length scales. \emph{Nature Communications} \textbf{2024}, \emph{15}, 3322\relax
\mciteBstWouldAddEndPuncttrue
\mciteSetBstMidEndSepPunct{\mcitedefaultmidpunct}
{\mcitedefaultendpunct}{\mcitedefaultseppunct}\relax
\EndOfBibitem
\bibitem[Ray \latin{et~al.}(2023)Ray, Mason, Boyens-Thiele, Farzadfard, Larsen, Norrild, Jahnke, and Buell]{Ray2023}
Ray,~S.; Mason,~T.~O.; Boyens-Thiele,~L.; Farzadfard,~A.; Larsen,~J.~A.; Norrild,~R.~K.; Jahnke,~N.; Buell,~A.~K. Mass photometric detection and quantification of nanoscale alpha-synuclein phase separation. \emph{Nature Chemistry} \textbf{2023}, \emph{15}, 1306--1316\relax
\mciteBstWouldAddEndPuncttrue
\mciteSetBstMidEndSepPunct{\mcitedefaultmidpunct}
{\mcitedefaultendpunct}{\mcitedefaultseppunct}\relax
\EndOfBibitem
\bibitem[Toledo \latin{et~al.}(2023)Toledo, Gianotti, Vazquez, and Ermácora]{Toledo2023}
Toledo,~P.~L.; Gianotti,~A.~R.; Vazquez,~D.~S.; Ermácora,~M.~R. Protein nanocondensates: the next frontier. \emph{Biophysical Reviews} \textbf{2023}, \emph{15}, 515--530\relax
\mciteBstWouldAddEndPuncttrue
\mciteSetBstMidEndSepPunct{\mcitedefaultmidpunct}
{\mcitedefaultendpunct}{\mcitedefaultseppunct}\relax
\EndOfBibitem
\bibitem[Jain \latin{et~al.}(2016)Jain, Wheeler, Walters, Agrawal, Barsic, and Parker]{Jain2016}
Jain,~S.; Wheeler,~J.; Walters,~R.; Agrawal,~A.; Barsic,~A.; Parker,~R. ATPase-Modulated Stress Granules Contain a Diverse Proteome and Substructure. \emph{Cell} \textbf{2016}, \emph{164}, 487--498\relax
\mciteBstWouldAddEndPuncttrue
\mciteSetBstMidEndSepPunct{\mcitedefaultmidpunct}
{\mcitedefaultendpunct}{\mcitedefaultseppunct}\relax
\EndOfBibitem
\bibitem[Folkmann \latin{et~al.}(2021)Folkmann, Putnam, Lee, and Seydoux]{Folkmann2021}
Folkmann,~A.~W.; Putnam,~A.; Lee,~C.~F.; Seydoux,~G. Regulation of biomolecular condensates by interfacial protein clusters. \emph{Science} \textbf{2021}, \emph{373}, 1218--1224\relax
\mciteBstWouldAddEndPuncttrue
\mciteSetBstMidEndSepPunct{\mcitedefaultmidpunct}
{\mcitedefaultendpunct}{\mcitedefaultseppunct}\relax
\EndOfBibitem
\bibitem[Rauscher and Pomès(2017)Rauscher, and Pomès]{Rauscher2017}
Rauscher,~S.; Pomès,~R. The liquid structure of elastin. \emph{eLife} \textbf{2017}, \emph{6}, e26526\relax
\mciteBstWouldAddEndPuncttrue
\mciteSetBstMidEndSepPunct{\mcitedefaultmidpunct}
{\mcitedefaultendpunct}{\mcitedefaultseppunct}\relax
\EndOfBibitem
\bibitem[Souza \latin{et~al.}(2021)Souza, Alessandri, Barnoud, Thallmair, Faustino, Gr{\"u}newald, Patmanidis, Abdizadeh, Bruininks, Wassenaar, Kroon, Melcr, Nieto, Corradi, Khan, Domański, Javanainen, Martinez-Seara, Reuter, Best, Vattulainen, Monticelli, Periole, Tieleman, de~Vries, and Marrink]{Souza2021}
Souza,~P. \latin{et~al.}  Martini 3: a general purpose force field for coarse-grained molecular dynamics. \emph{Nat. Methods} \textbf{2021}, \emph{10}, 382--388\relax
\mciteBstWouldAddEndPuncttrue
\mciteSetBstMidEndSepPunct{\mcitedefaultmidpunct}
{\mcitedefaultendpunct}{\mcitedefaultseppunct}\relax
\EndOfBibitem
\bibitem[Sami and Marrink(2023)Sami, and Marrink]{Sami2023}
Sami,~S.; Marrink,~S.~J. Reactive Martini: Chemical Reactions in Coarse-Grained Molecular Dynamics Simulations. \emph{J. Chem. Theory Comput.} \textbf{2023}, \emph{19}, 4040--4046\relax
\mciteBstWouldAddEndPuncttrue
\mciteSetBstMidEndSepPunct{\mcitedefaultmidpunct}
{\mcitedefaultendpunct}{\mcitedefaultseppunct}\relax
\EndOfBibitem
\bibitem[Pezeshkian \latin{et~al.}(2023)Pezeshkian, Gr{\"u}newald, Narykov, Lu, Arkhipova, Solodovnikov, Wassenaar, Marrink, and Korkin]{Pezeshkian2023}
Pezeshkian,~W.; Gr{\"u}newald,~F.; Narykov,~O.; Lu,~S.; Arkhipova,~V.; Solodovnikov,~A.; Wassenaar,~T.~A.; Marrink,~S.~J.; Korkin,~D. Molecular architecture and dynamics of SARS-CoV-2 envelope by integrative modeling. \emph{Structure} \textbf{2023}, \emph{31}, 492--503.e7\relax
\mciteBstWouldAddEndPuncttrue
\mciteSetBstMidEndSepPunct{\mcitedefaultmidpunct}
{\mcitedefaultendpunct}{\mcitedefaultseppunct}\relax
\EndOfBibitem
\bibitem[Vainikka and Marrink(2023)Vainikka, and Marrink]{Vainikka2023}
Vainikka,~P.; Marrink,~S.~J. Martini 3 Coarse-Grained Model for Second-Generation Unidirectional Molecular Motors and Switches. \emph{J. Chem. Theory Comput.} \textbf{2023}, \emph{19}, 596--604\relax
\mciteBstWouldAddEndPuncttrue
\mciteSetBstMidEndSepPunct{\mcitedefaultmidpunct}
{\mcitedefaultendpunct}{\mcitedefaultseppunct}\relax
\EndOfBibitem
\bibitem[Marrink \latin{et~al.}(2023)Marrink, Monticelli, Melo, Alessandri, Tieleman, and Souza]{Marrink2023}
Marrink,~S.~J.; Monticelli,~L.; Melo,~M.~N.; Alessandri,~R.; Tieleman,~D.~P.; Souza,~P. C.~T. Two decades of Martini: Better beads, broader scope. \emph{WIREs Computational Molecular Science} \textbf{2023}, \emph{13}, e1620\relax
\mciteBstWouldAddEndPuncttrue
\mciteSetBstMidEndSepPunct{\mcitedefaultmidpunct}
{\mcitedefaultendpunct}{\mcitedefaultseppunct}\relax
\EndOfBibitem
\bibitem[Abraham \latin{et~al.}(2015)Abraham, Murtola, Schulz, P{\'a}ll, Smith, Hess, and Lindahl]{Gromacs2015}
Abraham,~M.~J.; Murtola,~T.; Schulz,~R.; P{\'a}ll,~S.; Smith,~J.~C.; Hess,~B.; Lindahl,~E. \uppercase{GROMACS}: High performance molecular simulations through multi-level parallelism from laptops to supercomputers. \emph{SoftwareX} \textbf{2015}, \emph{1-2}, 19--25\relax
\mciteBstWouldAddEndPuncttrue
\mciteSetBstMidEndSepPunct{\mcitedefaultmidpunct}
{\mcitedefaultendpunct}{\mcitedefaultseppunct}\relax
\EndOfBibitem
\bibitem[Ing\'olfsson \latin{et~al.}(2023)Ing\'olfsson, Rizuan, Liu, Mohanty, Souza, Marrink, Bowers, Mittal, and Berry]{Ingolfsson2023}
Ing\'olfsson,~H.~I.; Rizuan,~A.; Liu,~X.; Mohanty,~P.; Souza,~P.~C.; Marrink,~S.~J.; Bowers,~M.~T.; Mittal,~J.; Berry,~J. Multiscale simulations reveal TDP-43 molecular-level interactions driving condensation. \emph{Biophysical Journal} \textbf{2023}, \emph{122}, 4370--4381\relax
\mciteBstWouldAddEndPuncttrue
\mciteSetBstMidEndSepPunct{\mcitedefaultmidpunct}
{\mcitedefaultendpunct}{\mcitedefaultseppunct}\relax
\EndOfBibitem
\bibitem[Thomasen \latin{et~al.}(2022)Thomasen, Pesce, Roesgaard, Tesei, and Lindorff-Larsen]{Thomasen2022}
Thomasen,~F.~E.; Pesce,~F.; Roesgaard,~M.~A.; Tesei,~G.; Lindorff-Larsen,~K. Improving Martini 3 for Disordered and Multidomain Proteins. \emph{J. Chem. Theory Comput.} \textbf{2022}, \emph{18}, 2033--2041\relax
\mciteBstWouldAddEndPuncttrue
\mciteSetBstMidEndSepPunct{\mcitedefaultmidpunct}
{\mcitedefaultendpunct}{\mcitedefaultseppunct}\relax
\EndOfBibitem
\bibitem[Kroon \latin{et~al.}(2023)Kroon, Gr{\"u}newald, Barnoud, van Tilburg, Wassenaar, and J.]{Kroon2023}
Kroon,~P.~C.; Gr{\"u}newald,~F.; Barnoud,~J.; van Tilburg,~P. A.~M.; Wassenaar,~P. C. S. T.~A.; J.,~S. Martinize2 and Vermouth: Unified Framework for Topology Generation. \emph{eLife} \textbf{2023}, \emph{12}, RP90627\relax
\mciteBstWouldAddEndPuncttrue
\mciteSetBstMidEndSepPunct{\mcitedefaultmidpunct}
{\mcitedefaultendpunct}{\mcitedefaultseppunct}\relax
\EndOfBibitem
\bibitem[MeloLab(2024)]{MeloLab_github}
MeloLab ATP parameters in Martini 3. 2024; https://github.com/MeloLab/ CofactorParameterization\relax
\mciteBstWouldAddEndPuncttrue
\mciteSetBstMidEndSepPunct{\mcitedefaultmidpunct}
{\mcitedefaultendpunct}{\mcitedefaultseppunct}\relax
\EndOfBibitem
\bibitem[Bussi \latin{et~al.}(2007)Bussi, Donadio, and Parrinello]{parinello}
Bussi,~G.; Donadio,~D.; Parrinello,~M. Canonical sampling through velocity rescaling. \emph{J. Chem. Phys.} \textbf{2007}, \emph{126}, 014101\relax
\mciteBstWouldAddEndPuncttrue
\mciteSetBstMidEndSepPunct{\mcitedefaultmidpunct}
{\mcitedefaultendpunct}{\mcitedefaultseppunct}\relax
\EndOfBibitem
\bibitem[Parrinello and Rahman(1981)Parrinello, and Rahman]{parinelloRahman}
Parrinello,~M.; Rahman,~A. Polymorphic transitions in single crystals: a new molecular dynamics method. \emph{J. Appl. Phys.} \textbf{1981}, \emph{52}, 7182--7190\relax
\mciteBstWouldAddEndPuncttrue
\mciteSetBstMidEndSepPunct{\mcitedefaultmidpunct}
{\mcitedefaultendpunct}{\mcitedefaultseppunct}\relax
\EndOfBibitem
\bibitem[de~Jong \latin{et~al.}(2016)de~Jong, Baoukina, Ing{\'o}lfsson, and Marrink]{martiniCutoff}
de~Jong,~D.~H.; Baoukina,~S.; Ing{\'o}lfsson,~H.~I.; Marrink,~S.~J. Martini straight: boosting performance using a shorter cut-off and \uppercase{GPU}s. \emph{Comp. Phys. Comm.} \textbf{2016}, \emph{199}, 1--7\relax
\mciteBstWouldAddEndPuncttrue
\mciteSetBstMidEndSepPunct{\mcitedefaultmidpunct}
{\mcitedefaultendpunct}{\mcitedefaultseppunct}\relax
\EndOfBibitem
\bibitem[Humphrey \latin{et~al.}(1996)Humphrey, Dalke, and Schulten]{VMD}
Humphrey,~W.; Dalke,~A.; Schulten,~K. \uppercase{VMD}: Visual molecular dynamics. \emph{J. Mol. Graph.} \textbf{1996}, \emph{14}, 33--38\relax
\mciteBstWouldAddEndPuncttrue
\mciteSetBstMidEndSepPunct{\mcitedefaultmidpunct}
{\mcitedefaultendpunct}{\mcitedefaultseppunct}\relax
\EndOfBibitem
\bibitem[Michaud-Agrawal \latin{et~al.}(2011)Michaud-Agrawal, Denning, Woolf, and Beckstein]{MDAnalysis2011}
Michaud-Agrawal,~N.; Denning,~E.~J.; Woolf,~T.~B.; Beckstein,~O. MDAnalysis: A toolkit for the analysis of molecular dynamics simulations. \emph{Journal of Computational Chemistry} \textbf{2011}, \emph{32}, 2319--2327\relax
\mciteBstWouldAddEndPuncttrue
\mciteSetBstMidEndSepPunct{\mcitedefaultmidpunct}
{\mcitedefaultendpunct}{\mcitedefaultseppunct}\relax
\EndOfBibitem
\bibitem[{R}ichard {J}.~{G}owers \latin{et~al.}(2016){R}ichard {J}.~{G}owers, {M}ax {L}inke, {J}onathan {B}arnoud, {T}yler {J}. {E}.~{R}eddy, {M}anuel {N}.~{M}elo, {S}ean {L}.~{S}eyler, {J}an {D}omański, {D}avid {L}.~{D}otson, {S}ébastien {B}uchoux, {I}an {M}.~{K}enney, and {O}liver {B}eckstein]{MDAnalysis2016}
{R}ichard {J}.~{G}owers; {M}ax {L}inke; {J}onathan {B}arnoud; {T}yler {J}. {E}.~{R}eddy; {M}anuel {N}.~{M}elo; {S}ean {L}.~{S}eyler; {J}an {D}omański; {D}avid {L}.~{D}otson; {S}ébastien {B}uchoux; {I}an {M}.~{K}enney; {O}liver {B}eckstein {M}{D}{A}nalysis: {A} {P}ython {P}ackage for the {R}apid {A}nalysis of {M}olecular {D}ynamics {S}imulations. {P}roceedings of the 15th {P}ython in {S}cience {C}onference. 2016; pp 98 -- 105\relax
\mciteBstWouldAddEndPuncttrue
\mciteSetBstMidEndSepPunct{\mcitedefaultmidpunct}
{\mcitedefaultendpunct}{\mcitedefaultseppunct}\relax
\EndOfBibitem
\bibitem[Wassenaar \latin{et~al.}(2014)Wassenaar, Pluhackova, B{\"o}ckmann, Marrink, and Tieleman]{Wassenaar2014}
Wassenaar,~T.~A.; Pluhackova,~K.; B{\"o}ckmann,~R.~A.; Marrink,~S.~J.; Tieleman,~D.~P. Going Backward: A Flexible Geometric Approach to Reverse Transformation from Coarse Grained to Atomistic Models. \emph{J. Chem. Theory Comput.} \textbf{2014}, \emph{10}, 676--690\relax
\mciteBstWouldAddEndPuncttrue
\mciteSetBstMidEndSepPunct{\mcitedefaultmidpunct}
{\mcitedefaultendpunct}{\mcitedefaultseppunct}\relax
\EndOfBibitem
\bibitem[Huang and MacKerell~Jr(2013)Huang, and MacKerell~Jr]{Huang2013}
Huang,~J.; MacKerell~Jr,~A.~D. CHARMM36 all-atom additive protein force field: Validation based on comparison to NMR data. \emph{Journal of Computational Chemistry} \textbf{2013}, \emph{34}, 2135--2145\relax
\mciteBstWouldAddEndPuncttrue
\mciteSetBstMidEndSepPunct{\mcitedefaultmidpunct}
{\mcitedefaultendpunct}{\mcitedefaultseppunct}\relax
\EndOfBibitem
\bibitem[Baker \latin{et~al.}(2001)Baker, Sept, Joseph, Holst, and McCammon]{Baker2001}
Baker,~N.~A.; Sept,~D.; Joseph,~S.; Holst,~M.~J.; McCammon,~J.~A. Electrostatics of nanosystems: Application to microtubules and the ribosome. \emph{Proceedings of the National Academy of Sciences} \textbf{2001}, \emph{98}, 10037--10041\relax
\mciteBstWouldAddEndPuncttrue
\mciteSetBstMidEndSepPunct{\mcitedefaultmidpunct}
{\mcitedefaultendpunct}{\mcitedefaultseppunct}\relax
\EndOfBibitem
\bibitem[Jurrus \latin{et~al.}(2018)Jurrus, Engel, Star, Monson, Brandi, Felberg, Brookes, Wilson, Chen, Liles, Chun, Li, Gohara, Dolinsky, Konecny, Koes, Nielsen, Head-Gordon, Geng, Krasny, Wei, Holst, McCammon, and Baker]{Jurrus2018}
Jurrus,~E. \latin{et~al.}  Improvements to the APBS biomolecular solvation software suite. \emph{Protein Science} \textbf{2018}, \emph{27}, 112--128\relax
\mciteBstWouldAddEndPuncttrue
\mciteSetBstMidEndSepPunct{\mcitedefaultmidpunct}
{\mcitedefaultendpunct}{\mcitedefaultseppunct}\relax
\EndOfBibitem
\bibitem[Chen \latin{et~al.}(2022)Chen, Yu, Yousefi, Iwahara, and Pettitt]{Chen2022}
Chen,~C.; Yu,~B.; Yousefi,~R.; Iwahara,~J.; Pettitt,~B.~M. Assessment of the Components of the Electrostatic Potential of Proteins in Solution: Comparing Experiment and Theory. \emph{J. Phys. Chem. B} \textbf{2022}, \emph{126}, 4543--4554\relax
\mciteBstWouldAddEndPuncttrue
\mciteSetBstMidEndSepPunct{\mcitedefaultmidpunct}
{\mcitedefaultendpunct}{\mcitedefaultseppunct}\relax
\EndOfBibitem
\bibitem[Kausik \latin{et~al.}(2010)Kausik, Srivastava, Korevaar, Stucky, Waite, and Han]{Kausik2010}
Kausik,~R.; Srivastava,~A.; Korevaar,~P.~A.; Stucky,~G.; Waite,~J.~H.; Han,~S. Local water dynamics in coacervated polyelectrolytes monitored through dynamic nuclear polarization-enhanced \uppercase{1H NMR}. \emph{Macromolecules} \textbf{2010}, \emph{43}, 3122\relax
\mciteBstWouldAddEndPuncttrue
\mciteSetBstMidEndSepPunct{\mcitedefaultmidpunct}
{\mcitedefaultendpunct}{\mcitedefaultseppunct}\relax
\EndOfBibitem
\bibitem[Mondal and Shakhnovich(2025)Mondal, and Shakhnovich]{Mondal2025}
Mondal,~S.; Shakhnovich,~E. The Origin of the Ionic-strength Dependent Reentrant Behavior in Liquid-Liquid Phase Separation of Neutral IDPs. \emph{bioRxiv} \textbf{2025}, 2025.03.20.644249\relax
\mciteBstWouldAddEndPuncttrue
\mciteSetBstMidEndSepPunct{\mcitedefaultmidpunct}
{\mcitedefaultendpunct}{\mcitedefaultseppunct}\relax
\EndOfBibitem
\bibitem[Wong \latin{et~al.}(2020)Wong, Kim, Muhandiram, Forman-Kay, and Kay]{Wong2020}
Wong,~L.~E.; Kim,~T.~H.; Muhandiram,~D.~R.; Forman-Kay,~J.~D.; Kay,~L.~E. NMR Experiments for Studies of Dilute and Condensed Protein Phases: Application to the Phase-Separating Protein CAPRIN1. \emph{Journal of the American Chemical Society} \textbf{2020}, \emph{142}, 2471--2489\relax
\mciteBstWouldAddEndPuncttrue
\mciteSetBstMidEndSepPunct{\mcitedefaultmidpunct}
{\mcitedefaultendpunct}{\mcitedefaultseppunct}\relax
\EndOfBibitem
\bibitem[LaCour \latin{et~al.}(2025)LaCour, Heindel, Zhao, and Head-Gordon]{LaCour2025}
LaCour,~R.~A.; Heindel,~J.~P.; Zhao,~R.; Head-Gordon,~T. The Role of Interfaces and Charge for Chemical Reactivity in Microdroplets. \emph{Journal of the American Chemical Society} \textbf{2025}, \emph{147}, 6299--6317\relax
\mciteBstWouldAddEndPuncttrue
\mciteSetBstMidEndSepPunct{\mcitedefaultmidpunct}
{\mcitedefaultendpunct}{\mcitedefaultseppunct}\relax
\EndOfBibitem
\bibitem[Anderson(1989)]{anderson1989colloid}
Anderson,~J.~L. Colloid transport by interfacial forces. \emph{Annual Review of Fluid Mechanics} \textbf{1989}, \emph{21}, 61--99\relax
\mciteBstWouldAddEndPuncttrue
\mciteSetBstMidEndSepPunct{\mcitedefaultmidpunct}
{\mcitedefaultendpunct}{\mcitedefaultseppunct}\relax
\EndOfBibitem
\bibitem[Ahangama~Liyanage and Ditlev(2024)Ahangama~Liyanage, and Ditlev]{Liyanage2024}
Ahangama~Liyanage,~L.; Ditlev,~J.~A. Mesoscale condensates organize the cytoplasm. \emph{Nature Cell Biology} \textbf{2024}, \emph{26}, 310--312\relax
\mciteBstWouldAddEndPuncttrue
\mciteSetBstMidEndSepPunct{\mcitedefaultmidpunct}
{\mcitedefaultendpunct}{\mcitedefaultseppunct}\relax
\EndOfBibitem
\end{mcitethebibliography}
\bibliographystyle{achemso}

\newpage
\textbf{TOC Graphic}
\begin{figure}[h!]
\centering
\includegraphics[width=0.99\textwidth]{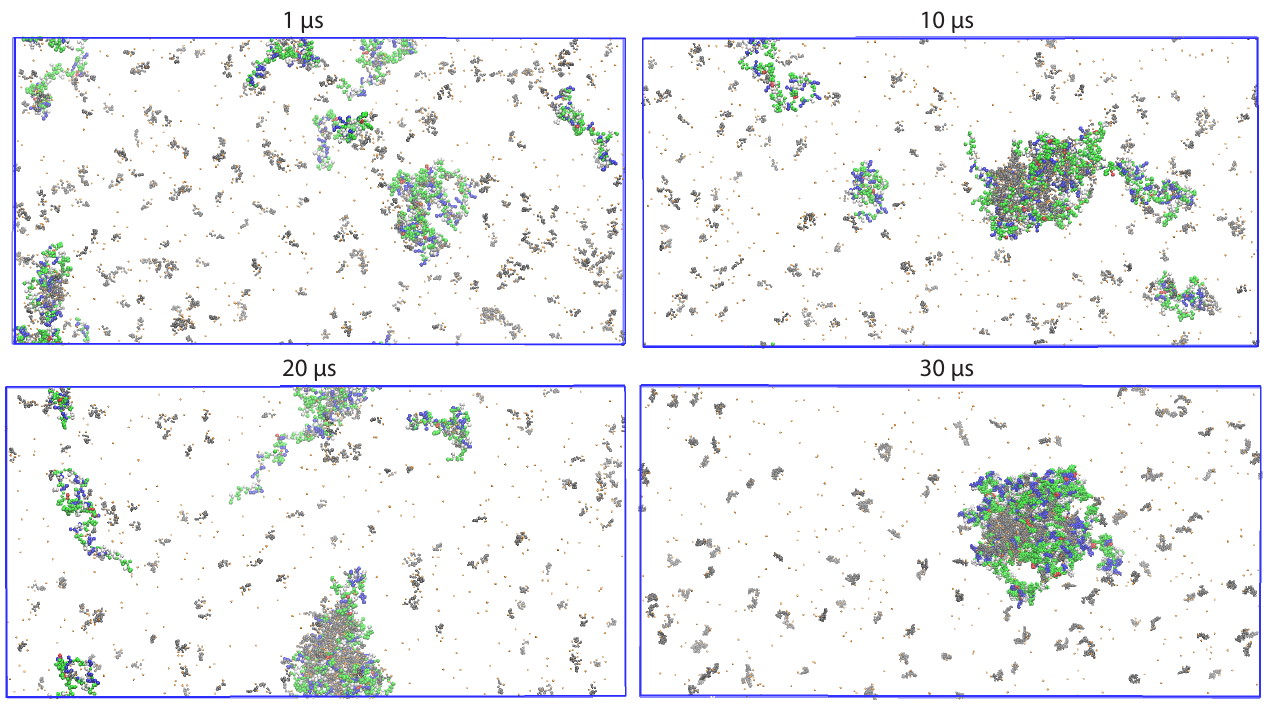}
\end{figure}
\end{document}


\begin{figure} [H]
\centering
\includegraphics[width=1.0\textwidth]{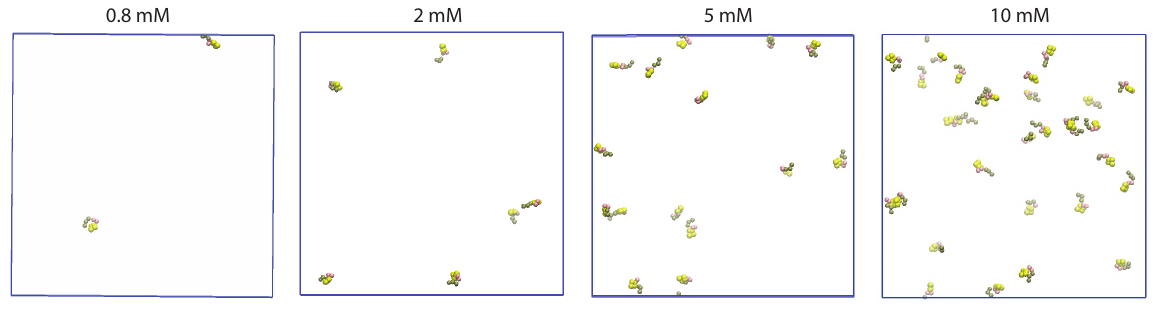}
\caption{\textit{ATP oligomerization without the Caprin1 protein.} No ATP oligomers were formed in the systems up to 5 mM, whereas for the system of 10 mM ATP concentration, only dimers were formed.}
\end{figure}

\begin{figure} [H]
\centering
\includegraphics[width=1.0\textwidth]{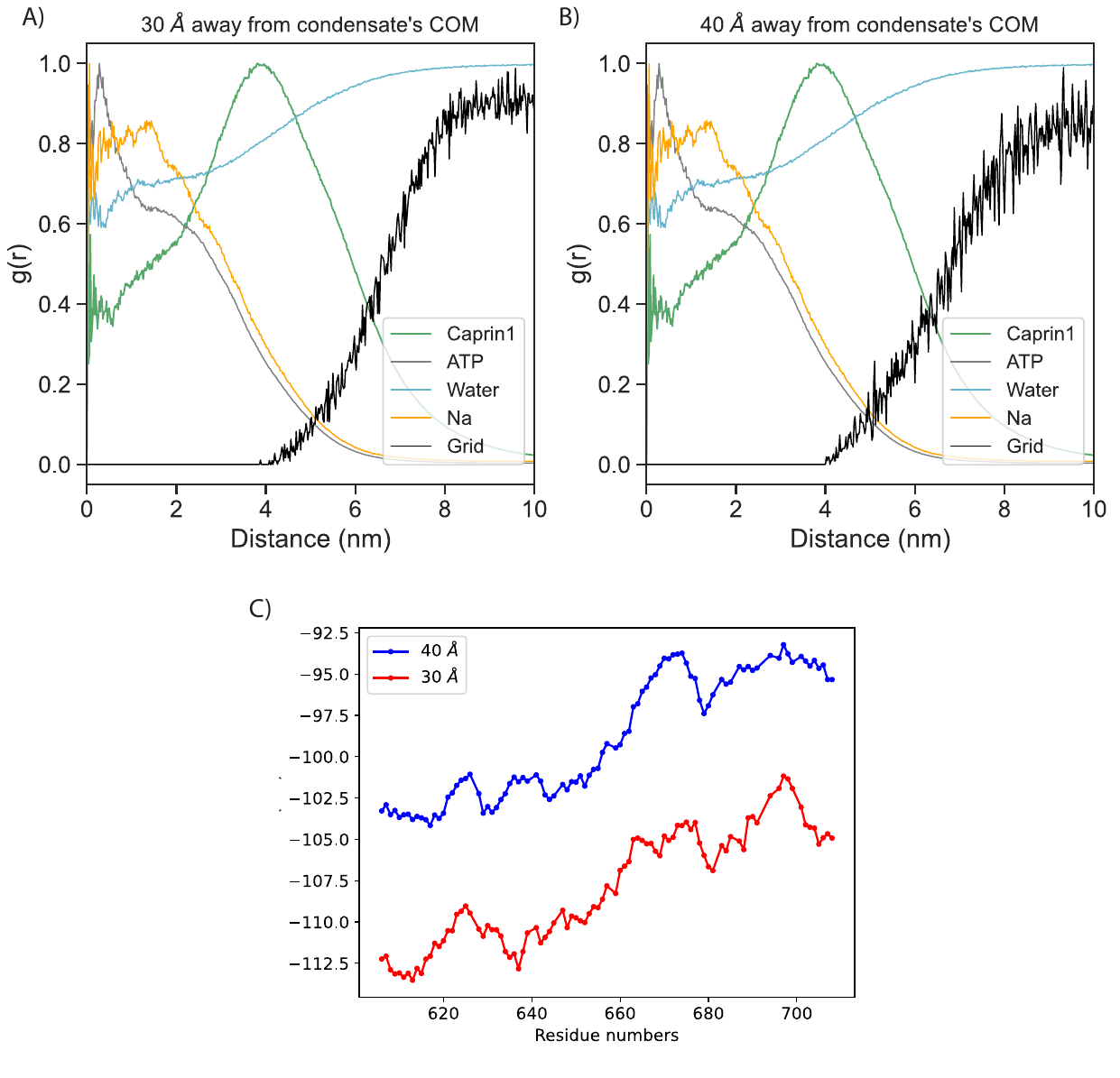}
\caption{\textit{Accessibility of the PRE probe and effect on NS-ESP for the condensate.} The density profile of all components of the Caprin1-ATP condensate shown in green for the Caprin1 chains, in gray for ATP, in cyan for water molecules, in orange for Na counterions and in black for the grid points. The density of probe sampling grid points is set at \textbf{(A)} 30 {\AA}, \textbf{(B)} 40 {\AA} away from the center of mass of the condensate.}
\end{figure}

\begin{figure} [H]
\centering
\includegraphics[width=1.0\textwidth]{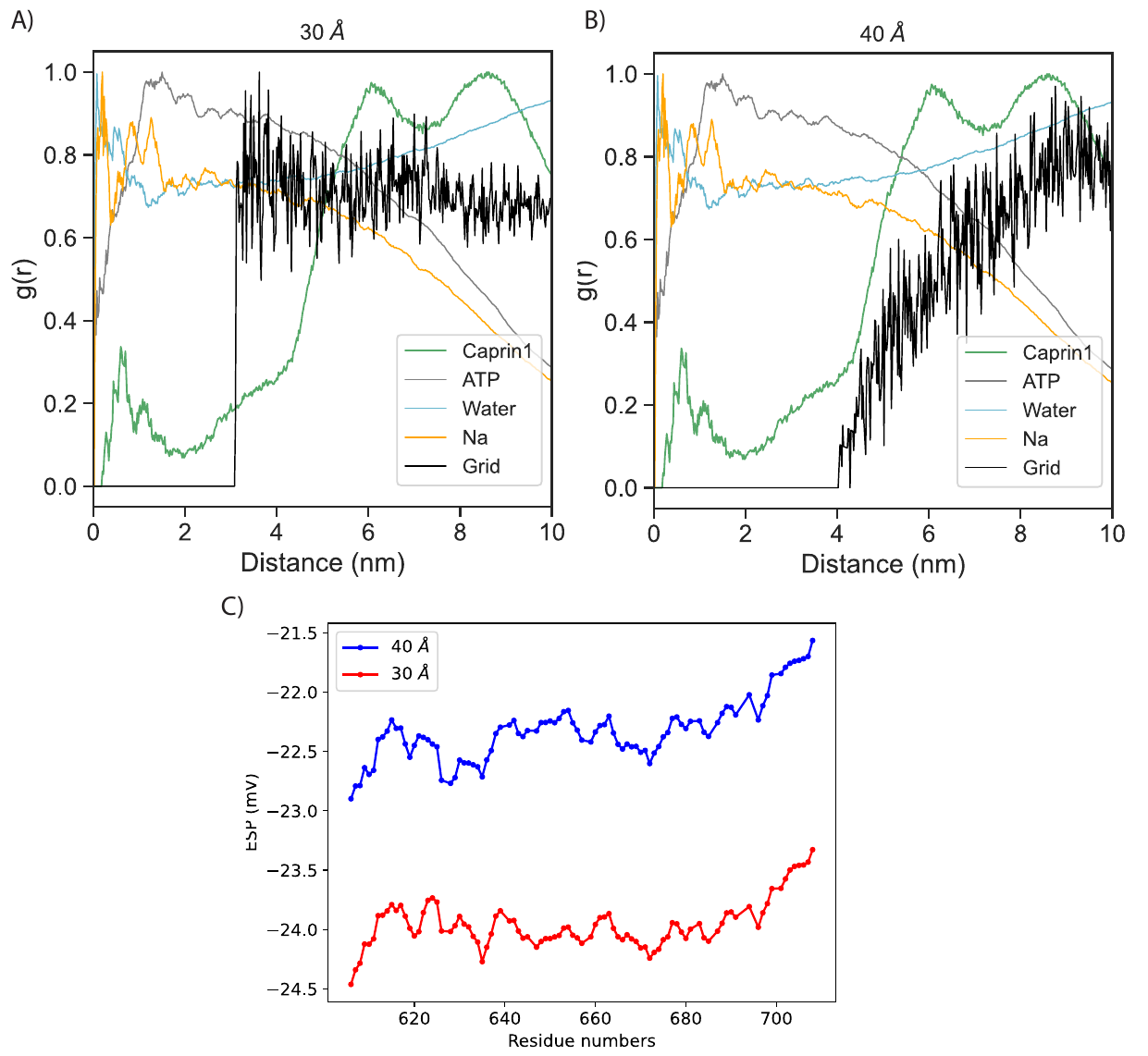}
\caption{\textit{Accessibility of the PRE probe and effect on NS-ESP for the broken condensate.} The density profile of all components of the Caprin1-ATP broken condensate shown in green for the Caprin1 chains, in gray for ATP, in cyan for water molecules, in orange for Na counterions and in black for the grid points. The density of probe sampling grid points is set at \textbf{(A)} 30 {\AA}, \textbf{(B)} 40 {\AA} away from the center of mass of the broken condensate.}
\end{figure}